\documentclass[twocolumn,showpacs,prb,superscriptaddress,a4paper,floatfix]{revtex4}
\usepackage{amssymb}
\usepackage{graphicx}
\usepackage{amsmath}
\usepackage{graphicx}
\usepackage{dcolumn}
\usepackage{bm}

\begin{document}

\title{Electronic Transport in Disordered Bilayer and Trilayer Graphene}
\author{Shengjun Yuan}
\email{s.yuan@science.ru.nl}
\affiliation{Institute for Molecules and Materials, Radboud University of Nijmegen,
NL-6525ED Nijmegen, The Netherlands}
\author{Hans De Raedt}
\affiliation{Department of Applied Physics, Zernike Institute for Advanced Materials,
University of Groningen, Nijenborgh 4, NL-9747AG Groningen, The Netherlands}
\author{Mikhail I. Katsnelson}
\affiliation{Institute for Molecules and Materials, Radboud University of Nijmegen,
NL-6525ED Nijmegen, The Netherlands}
\date{\today }

\begin{abstract}
We present a detailed numerical study of the electronic transport properties
of bilayer and trilayer graphene within a framework of single-electron
tight-binding model. Various types of disorder are considered, such as
resonant (hydrogen) impurities, vacancies, short- or long-range Gaussian
random potentials, and Gaussian random nearest neighbor hopping. The
algorithms are based on the numerical solution of the time-dependent Schr%
\"{o}dinger equation and applied to calculate the density of states and
conductivities (via the Kubo formula) of large samples containing millions
of atoms. In the cases under consideration, far enough from the neutrality
point, depending on the strength of disorders and the stacking sequence, a
linear or sublinear electron-density dependent conductivity is found. The
minimum conductivity $\sigma _{\min }\approx 2e^{2}/h$ (per layer) at the
charge neutrality point is the same for bilayer and trilayer graphene,
independent of the type of the impurities, but the plateau of minimum
conductivity around the neutrality point is only observed in the presence of
resonant impurities or vacancies, originating from the formation of the
impurity band.
\end{abstract}

\maketitle

\section{Introduction}

Graphene is a subject of numerous investigations motivated by its unique
electronic and lattice properties, interesting both conceptually and for
applications (for reviews, see Refs. %
\onlinecite{r1,r2,r3,r4,r5,r6,r7,Cresti2008,Mucciolo2010,Peres2010}). Single
layer graphene (SLG) is the two-dimensional crystalline form of carbon with
a linear electronic spectrum and chiral (A-B sublattice) symmetry, whose
extraordinary electron mobility and other unique features hold great promise
for nanoscale electronics and photonics. Bilayer and trilayer graphenes,
which are made out of two and three graphene planes, have also been produced
by the mechanical friction and motivated a lot of researches on their
transport properties \cite%
{natphys,falko2006,GCNT2006,Ohta2006,McCann2006,Koshino2006,Santos2007,
Katsnelson_bilayer,Castro2007,Castro2008,Nilsson2007,Morozov2008,Nilsson2008, Nilsson2006,Nilsson2006b, ZhuW2009,DasSarma2010,XiaoS2010,Lv2010,ZhangF2010, Zhu2010,Pereira2009,Ribeiro2008,Mallet2007,Bostwick2007,Gorbachev2007,LinYM2008,Nakamura2008,Feldman2009,Trushin2010}%
. The charge-carrying quasiparticles in bilayer graphene (BLG) obey
parabolic dispersion with non-zero mass, but retain a chiral nature similar
to that in SLG (with the Berry phase $2\pi $ instead of $\pi $) \cite%
{natphys,falko2006}. Furthermore, an electronic bandgap can be introduced in
a dual gate BLG \cite%
{McCann2006,Castro2007,Oostinga2008,ZhangYB2009,Castro2010,Taychatanapat2010}%
, and it makes BLG very appealing from the point of view of applications.
The trilayer graphene (TLG) is shown to have different electronic properties
which is strongly dependent on the interlayer stacking sequence \cite%
{Craciun2009,Koshino2009}. Nevertheless, graphene layers in real experiments
always have different kinds of disorder, such as ripples, adatoms,
admolecules, etc. One of the most important problems in the potential
applications of graphene in electronics, is understanding the effect of
these imperfections on the electronic structure and transport properties.

The scattering theory for Dirac electrons in SLG is discussed in Refs. %
\onlinecite{Shon1998,Peres2006,Katsnelson2007,Hentschel2007,Novikov2007}.
Long-range scattering centers are of special importance for transport
properties of SLG, such as charge impurities \cite%
{r6,Nomura2006,Ando2006,Hwang2007,Adama2009}, ripples created long-range
elastic deformations \cite{r7,Katsnelson2008}, and resonant scattering
centers \cite%
{Peres2006,Katsnelson2007,Katsnelson2008,Ostrovsky2006,Stauber2007,Titov2010,Robinson2008,Wehling2010,Yuan2010,Ni2010}%
. Recently, the impact of charged impurity scattering on electronic
transport in BLG have been investigated theoretically \cite%
{Katsnelson_bilayer,DasSarma2010,Lv2010} and experimentally \cite{XiaoS2010}%
. The linear density-dependent conductivity at high density and the minimum
conductivity behavior around the charge neutrality point are expected \cite%
{Katsnelson_bilayer,DasSarma2010,Lv2010} and confirmed \cite{XiaoS2010}, but
the experimental results also suggest that charged impurity scattering alone
is not sufficient to explain the observed transport properties of pristine
BLG on SiO$_{2}$ before potassium doping \cite{XiaoS2010}. One possible
explanation of the experimental results might be the opening of a gap at the
Dirac point in biased BLG \cite{XiaoS2010}. On the other hand, some recent
experimental \cite{Ni2010} and theoretical \cite%
{Robinson2008,Wehling2010,Yuan2010} evidences appeared that the resonant
scattering due to carbon-carbon bonds between organic admolecules and
graphene (or by hydrogen impurities which are almost equivalent to C-C bonds
in a sense of electron scattering \cite{Wehling2010}) is the main
restricting factor for electron mobility in SLG on a substrate. These
results suggest that the resonant impurity could also be the dominant factor
of the transport properties of BLG and TLG.

\begin{figure*}[t]
\begin{center}
\includegraphics[width=10cm]{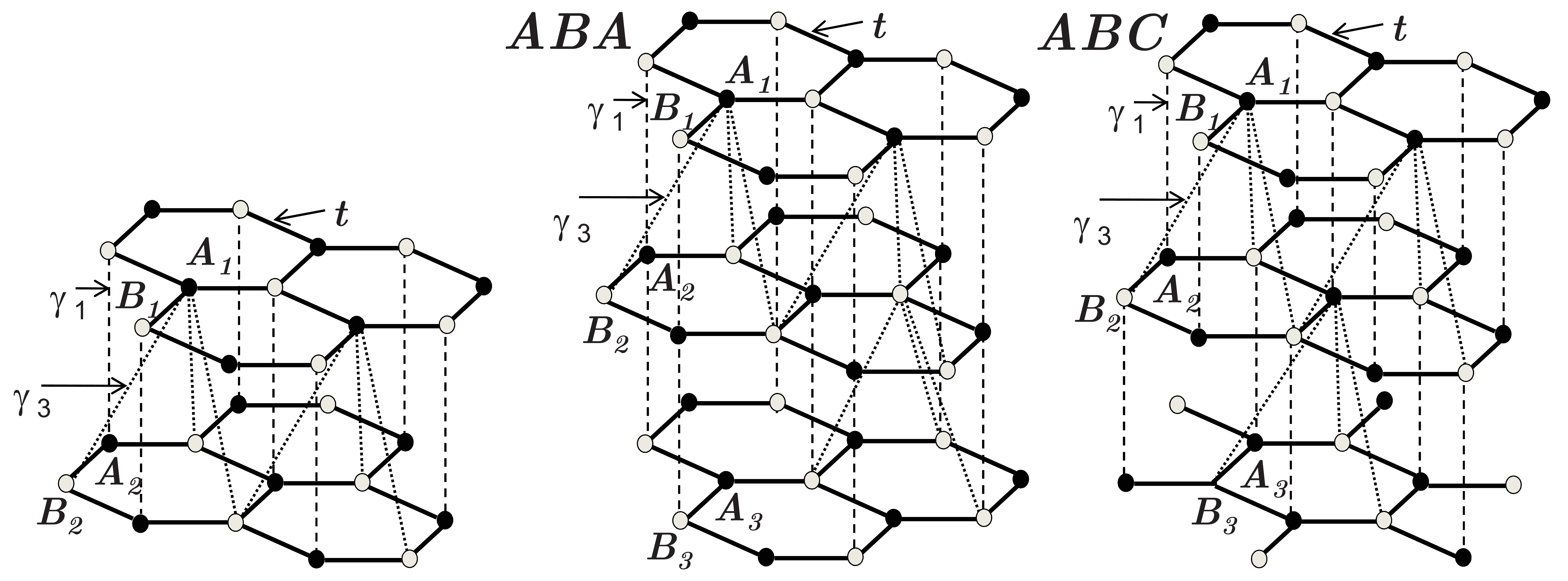}
\end{center}
\caption{Atomic structure of bilayer, ABA- and ABC-stacked trilayer
graphene. }
\label{multilayer3d}
\end{figure*}

In the present paper, we study the effect of different types of impurities
on the transport properties of graphene layers by direct numerical
simulations in a framework of the single-electron tight-binding model. We
consider four different types of defects: resonant (\textquotedblleft
hydrogen\textquotedblright ) impurities, vacancies, Gaussian on-site
potentials and Gaussian nearest carbon-carbon hoppings. The resonant
impurities/vacancies and the centers of the Gaussian potentials/couplings
are randomly introduced in the graphene layers. Our numerical calculations
are based on the time-evolution method \cite{Hams2000,DeRaedt2008,Yuan2010},
i.e., the time-evolution of the wave functions according to the Schr\"{o}%
dinger equation with additional averaging over a random superposition of
basis states. The main idea is that by performing Fourier transform of
various correlation functions, such as the wave function-wave function and
current-current correlation functions (Kubo formula), one can calculate the
electronic structure and transport properties such as the density of states
(DOS), quasieigenstates, ac (optical) and dc conductivities. The details of
the numerical method are presented in Ref. \onlinecite{Yuan2010}. The
advantages of the time-evolution method is that it allows us to carry out
calculations for very large systems, up to hundreds of millions of sites,
with a computational effort that increases only linearly with the system
size.

The paper is organized as follows. Section II gives a description of the
tight-binding Hamiltonian of multilayer graphene. In section III, IV, V and
VI, we focus on four different types of disorders respectively: resonant
impurities, vacancies, potential impurities, and nearest carbon-carbon
hopping impurities. Finally a brief discussion is given in section VII.

\section{Tight-binding model}

In general, the tight-binding Hamiltonian of multilayer graphene is given by
\begin{equation*}
H=\sum_{l=1}^{N_{layer}}H_{l}+\sum_{l=1}^{N_{layer}-1}H_{l}^{\prime },
\end{equation*}%
where $H_{l}$ is the Hamiltonian of SLG for $l$'th layer and $H_{l}^{\prime
} $ describes the hopping between layers $l$ and $l+1$.

The single-layer Hamiltonian $H_{l}$ is given by%
\begin{equation}
H_{l}=H_{0}+H_{v}+H_{imp},  \label{Hamiltonian}
\end{equation}%
where $H_{0}$ derives from the nearest neighbor hopping between the carbon
atoms:%
\begin{equation}
H_{0}=-\sum_{<i,j>}t_{ij}c_{i}^{+}c_{j},
\end{equation}%
$H_{v}$ denotes the on-site potential of the carbon atoms:%
\begin{equation}
H_{v}=\sum_{i}v_{i}c_{i}^{+}c_{i},
\end{equation}%
and $H_{imp}$ describes the resonant impurities (adatoms or admolecules):%
\begin{equation}
H_{imp}=\varepsilon _{d}\sum_{i}d_{i}^{+}d_{i}+V\sum_{i}\left(
d_{i}^{+}c_{i}+H.c.\right) .
\end{equation}

The interlayer Hamiltonian $H_{l}^{\prime }$ of bilayer graphene with AB
Bernal stacking is given by%
\begin{equation}
H_{l}^{\prime }=-\gamma _{1}\sum_{j}\left[ a_{l,j}^{+}b_{l+1,j}+H.c.\right]
-\gamma _{3}\sum_{j,j^{\prime }}\left[ b_{l,j}^{+}a_{l+1,j^{\prime }}+H.c.%
\right] ,
\end{equation}%
where $a_{m,i}^{+}$ ($b_{m,j}$) annihilates an electron on sublattice A (B),
in plane $m=l,l+1$, at site $R$ (see the atomic structure in Fig. \ref%
{multilayer3d}). Thus, the second layer in BLG is rotated with respect to
the first one by $+120^{\circ }$. For the third layer there are two options:
either the third carbon layer will be rotated with respect to the second
layer by $-120^{\circ }$ (than it will be exactly under the first layer) or
by $+120^{\circ }$. In the first case we have ABA-stacked trilayer graphene,
and in the second we have ABC-stacked (rhombohedral) graphene. The atomic
structures of the ABA- and ABC-stacked trilayer graphene are shown in Fig. %
\ref{multilayer3d}. These stacked sequences can be extended to multilayers,
i.e., the direct extension of ABA- and ABC-stacked sequences from trilayer
to quartic-layer are ABAB and ABCD. The spin degree of freedom contributes
only through a degeneracy factor and, for simplicity, is omitted in Eq.~(\ref%
{Hamiltonian}).

The density of states is obtained by Fourier transformation of the wave
function at time zero and time $t$:
\begin{equation}
\rho \left( \varepsilon \right) =\frac{1}{2\pi }\int_{-\infty }^{\infty
}e^{i\varepsilon t}\left\langle \varphi \left( 0\right) |\varphi \left(
t\right) \right\rangle dt,  \label{dosf}
\end{equation}%
where $\left\vert \varphi \left( 0\right) \right\rangle $ is an initial
random superposition state of all the basis states and $\left\vert \varphi
\left( t\right) \right\rangle =e^{-iHt}\left\vert \varphi \left( 0\right)
\right\rangle $ is calculated numerically according to the time-dependent
Schr\"{o}dinger equation (we use units with $\hbar =1$). A detailed
description of this method can be found in Refs. %
\onlinecite{Hams2000,Yuan2010}. The charge density is obtained by the
intergral of the density of states, i.e., $n_{e}\left( E\right)
=\int_{0}^{E}\rho \left( \varepsilon \right) d\varepsilon .$

The static (dc) conductivity is calculated by using the Kubo formula
\begin{equation}
\mathbf{\sigma }=-\frac{1}{A}Tr\left\{ \frac{\partial f}{\partial H}%
\int_{0}^{\infty }dt\frac{1}{2}\left[ JJ\left( t\right) +J\left( t\right) J%
\right] \right\} .  \label{conduc2}
\end{equation}%
where $J$ is the current operator and $A$ is the sample area. The main idea
of the calculation is to perform the time evolution of $\left\vert \varphi
\left( 0\right) \right\rangle $. Then, we can extract not only the DOS but
also the \textit{quasieigenstates} $\left\vert \Psi \left( \varepsilon
\right) \right\rangle $ \cite{Yuan2010}, which are superpositions of
degenerated energy-eigenstates. The conductivity at zero temperature can be
represented as
\begin{equation}
\mathbf{\sigma }=\frac{\rho \left( \varepsilon \right) }{V}\int_{0}^{\infty
}dt\text{Re}\left[ e^{-i\varepsilon t}\left\langle \varphi \left( 0\right)
\right\vert Je^{iHt}J\left\vert \varepsilon \right\rangle \right] ,
\label{conducappr}
\end{equation}%
where $\left\vert \varepsilon \right\rangle $ is defined as%
\begin{equation}
\left\vert \varepsilon \right\rangle =\frac{1}{\left\vert \left\langle
\varphi \left( 0\right) |\Psi \left( \varepsilon \right) \right\rangle
\right\vert }\left\vert \Psi \left( \varepsilon \right) \right\rangle .
\label{quasieigenstate}
\end{equation}

The accuracy of the numerical results is mainly determined by three
factors: the time interval of the propagation, the total number of time
steps, and the size of the sample.
In the numerical calculations, the integrals in Eq.~(\ref{dosf}) and
(\ref{conducappr}) are calculated using the Fast Fourier Transform (FFT).
According to the Nyquist sampling theorem, employing a sampling
interval $\Delta t=\pi /\max_{i}\left\vert E_{i}\right\vert $,
where $E_{i}$
are the eigenenergies, is sufficient to cover the
full range of eigenvalues.
In practice, we do not know
$\max_{i}\left\vert E_{i}\right\vert$ exactly
but it is easy to compute an upperbound (for instance the 1-norm of $H$)
such that $\Delta t$ can be considered as fixed.

In the present paper, the time evolution 
is calculated by the Chebyshev polynomial method, which has the same
accuracy as the machine's precision independent of the value of time
interval $\Delta t$.
Alternatively, one could use Suzuki's product formula decomposition
of the exponential operators for the tight-binding Hamiltonian~\cite{Kawarabayashi1995},
introducing another time step that has
to be (much) smaller than $\Delta t$ to obtain accurate results~\cite{Hams2000}.
In both cases, the accuracy of the energy eigenvalues is
determined by the total number of the propagation time steps $(N_{t})$
that is the number of the data items used in the FFT.
Eigenvalues that differ less than $\Delta E=\pi /N_{t}\Delta t$
cannot be identified properly. However, since
$\Delta E$ is proportional to $N_{t}^{-1}$
we only have to extend the length of the calculation
by a factor of two to increase the accuracy by the same factor.

The third factor which determines the accuracy of our numerical results is the
size of the sample. A sample with more sites in the real space will have more
random coefficients in the initial state
$\left\vert \varphi \left(0\right) \right\rangle$,
providing a better statistical representation of the superposition of all energy eigenstates.
This, however, is not a real issue in practice as it has be shown that
the statistical fluctuations vanish with the inverse of the dimension of the
Hilbert space~\cite{Hams2000}, which for our problem, is proportional to the number of sites
in the sample.
A comparison of the DOS calculated from different samples size was shown in Ref.~\onlinecite{Yuan2010},
which clearly shows that larger sample size leads to
better accuracy, and the result calculated from a SLG with $4096\times 4096$
lattice sites matches very well with the analytical expression~\cite{Yuan2010}.
More details on the numerical method itself can be found in Ref.~\onlinecite{Yuan2010}.
The values of conductivity presented in this paper
are normalized per layer and are expressed in units $e^{2}/h$.

Obviously, computer memory and CPU time evidently limit the size of the
graphene system that can be simulated. The required CPU time is mainly
determined by the number of operations to be performed on the state of the
system, but this imposes no hard limit. However, the memory of
the computer does. In the tight-binding approximation, a state
$\left\vert\varphi \right\rangle $ of a sample consisting by $N_{c}$
atoms is represented by a complex-valued vector of length $D=N_{c}$.
For numerical accuracy (and in view of the large number of arithmetic operators performed),
it is advisable to use $13-15$ digit floating-point arithmetic
(corresponding to $8$ bytes per real number).
Thus, to represent the state $\left\vert \varphi \right\rangle $ we need at least
$N_{c}\times 2^{4}$ bytes.
For example, for $N_{c}=4096\times 4096\sim 1.6\times 10^{7}$ we need $256$ MB of
memory to store a single arbitrary state
$\left\vert \varphi \right\rangle $.
This amount of memory is not a problem for the calculation of DOS
on a modest desktop PC or notebook, but it limits the calculation of the dc conductivity
on such machines.
To calculate one value of
$\sigma \left( \varepsilon\right) $
one needs storage of the corresponding quasieigenstate
$\left\vert \varepsilon \right\rangle $, and with
typically $64$ of such quasieigenstates in our simulations,
a sample of $N_{c}=4096\times 4096$ sites requires at least
$16$ GB memory for the storage, which is still
reasonable for present-day computer equipment.

\begin{figure}[t]
\begin{center}
\mbox{
\includegraphics[width=6.8cm]{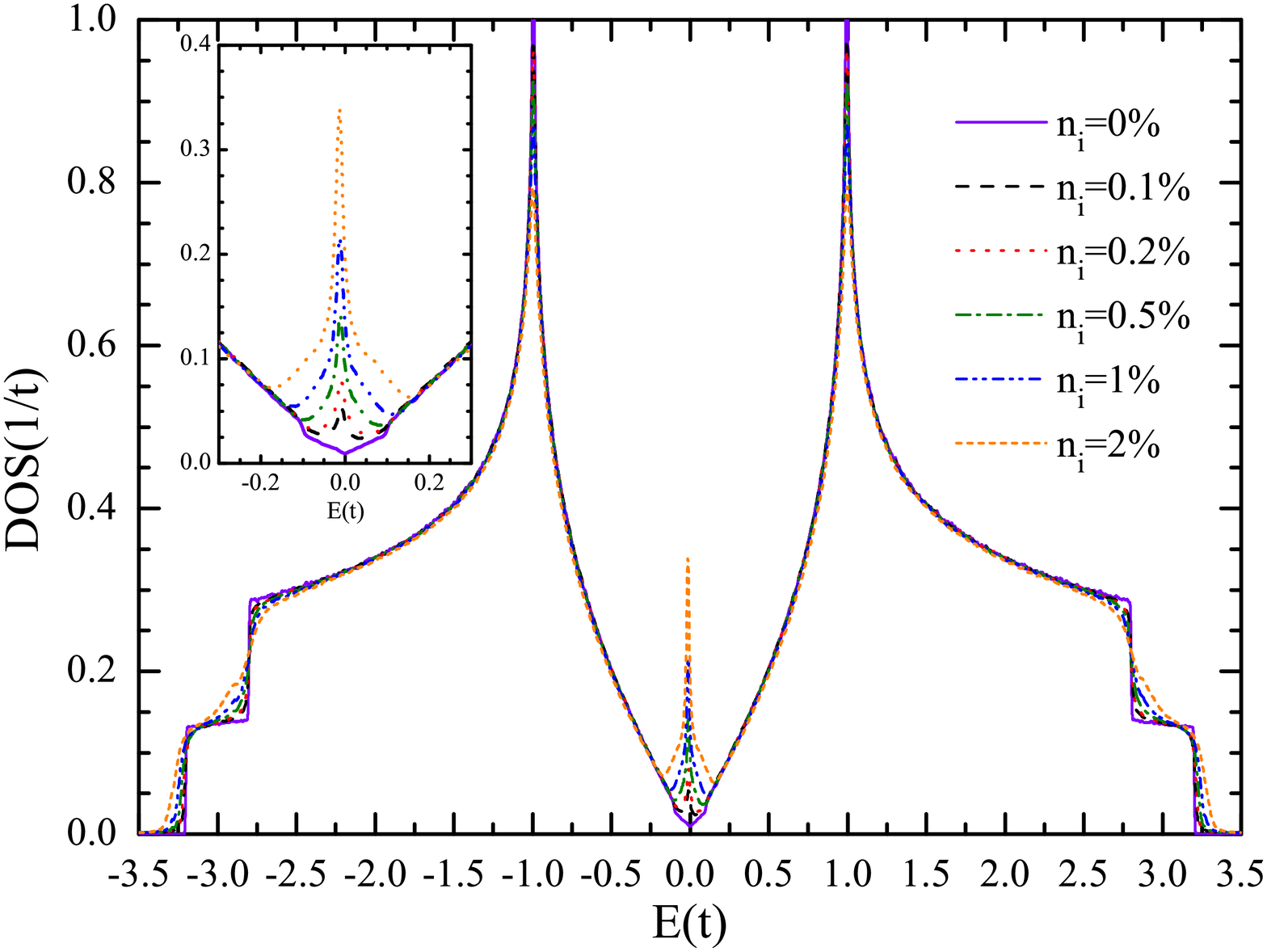}
} \mbox{
\includegraphics[width=6.8cm]{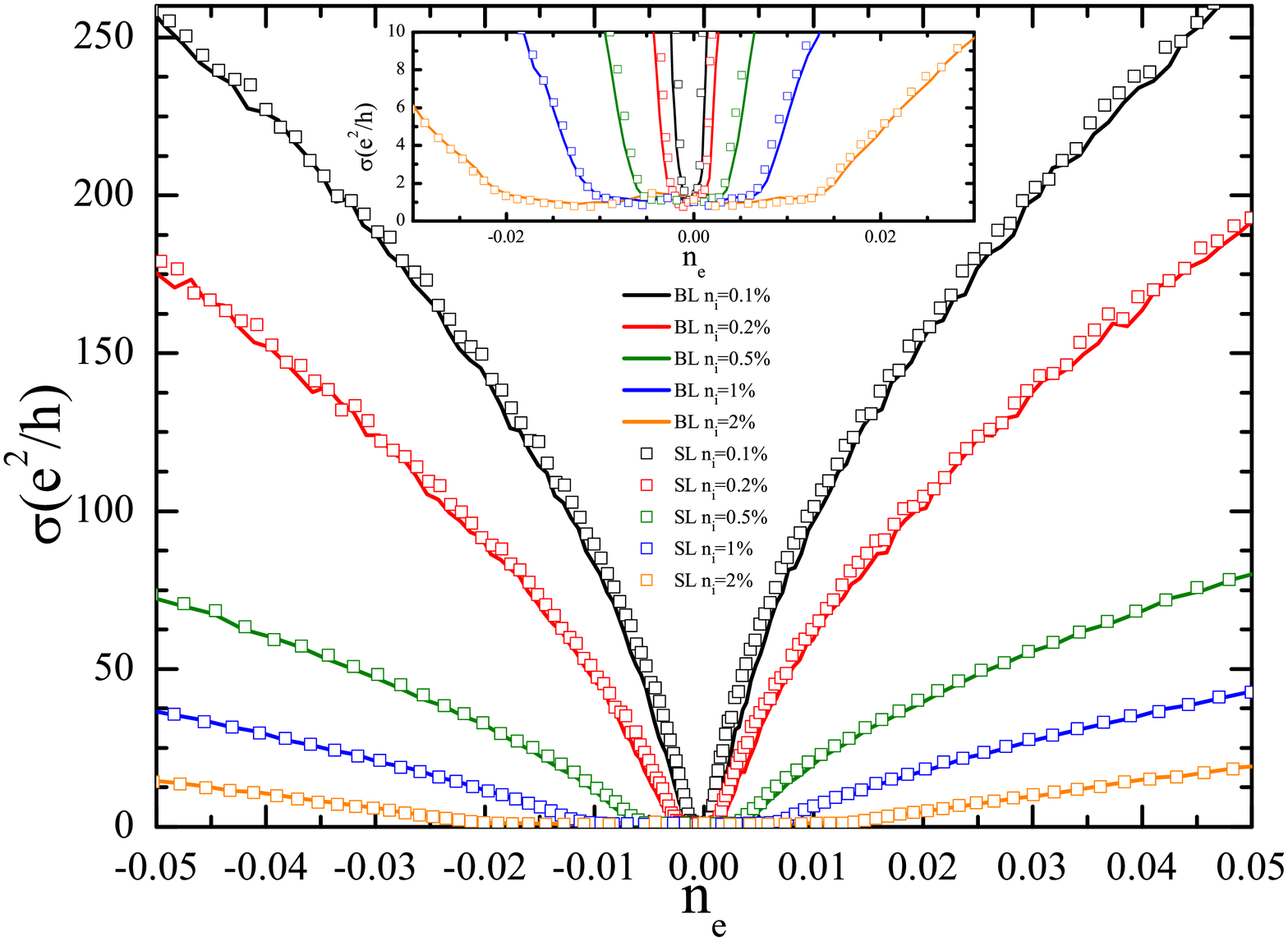}
} \mbox{
\includegraphics[width=6.8cm]{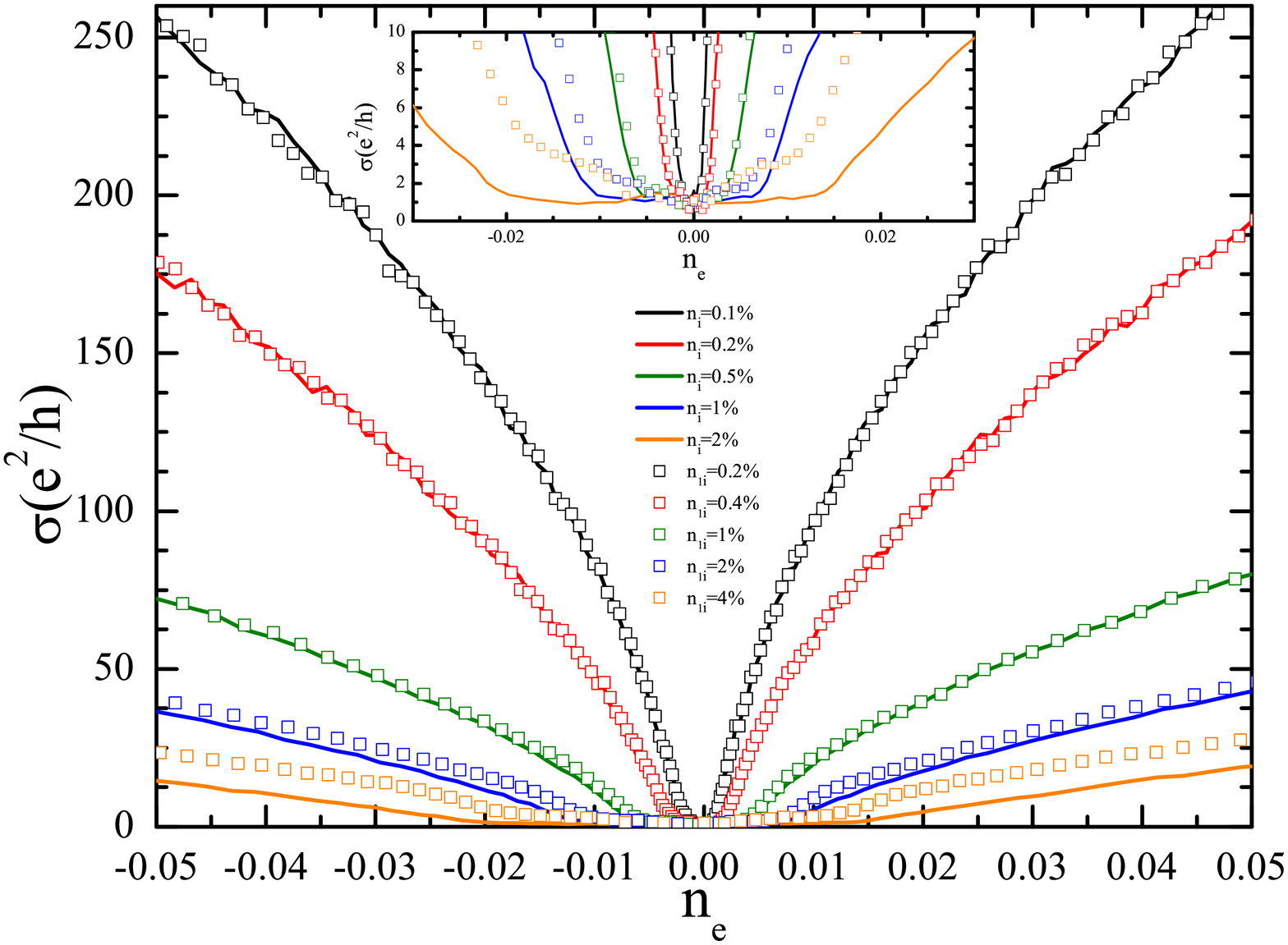}
}
\end{center}
\caption{(Colour online) Top panel: DOS of bilayer graphene ($\protect\gamma %
_{1}=\protect\gamma _{3}=0.1t$) with different concentrations of resonant
impurities ($\protect\varepsilon _{d}=-t/16,$ $V=2t$) added on both layers.
Middle panel: Comparison of the conductivity of the BLG (line) and SLG
(square) with the same concentration of resonant impurities. Bottom panel:
Comparison of the conductivity of bilayer graphene ($\protect\gamma _{1}=%
\protect\gamma _{3}=0.1t$) with the same amount of resonant impurities ($%
\protect\varepsilon _{d}=-t/16,$ $V=2t$) added on both layers (line, $n_{i}$%
) or only one layer (triangle, $n_{1i}$). Each layer in BLG contains $%
4096\times 4096$ carbon atoms, and SLG contains $6400\times 6400\,$carbon
atoms.}
\label{ximp}
\end{figure}

\begin{figure*}[t]
\begin{center}
\mbox{
\includegraphics[width=7.5cm]{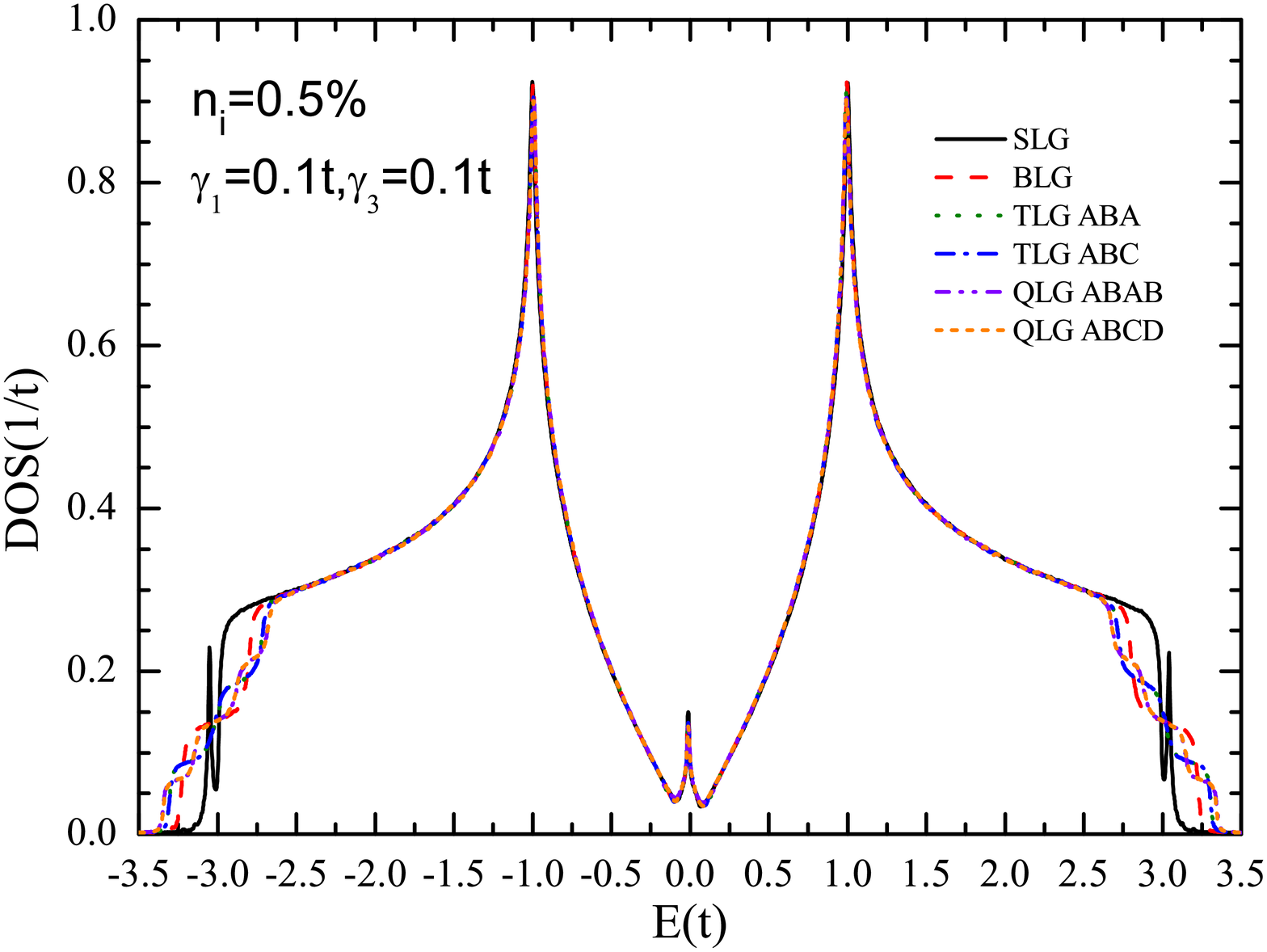}
\includegraphics[width=7.5cm]{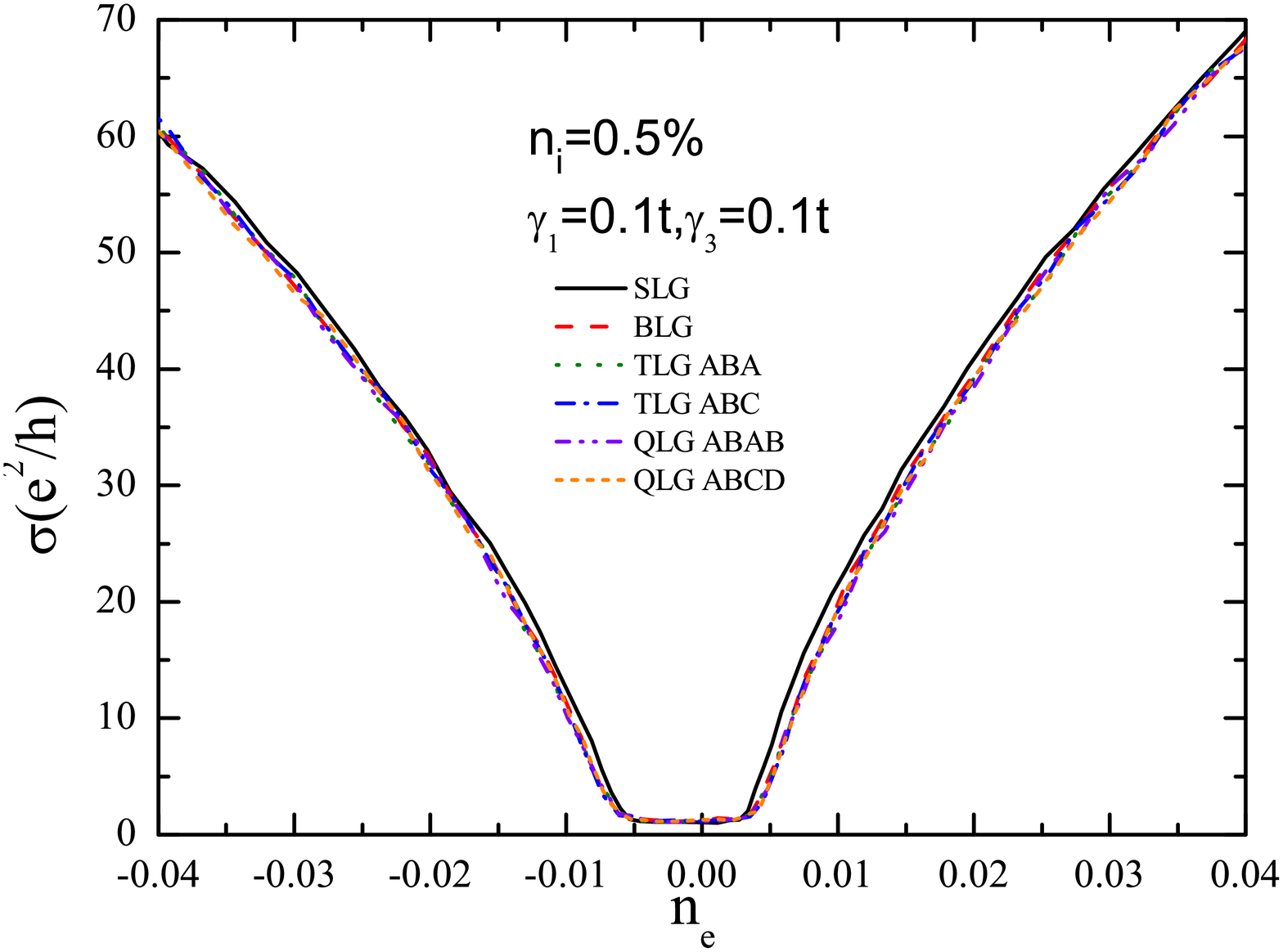}
}
\mbox{
\includegraphics[width=7.5cm]{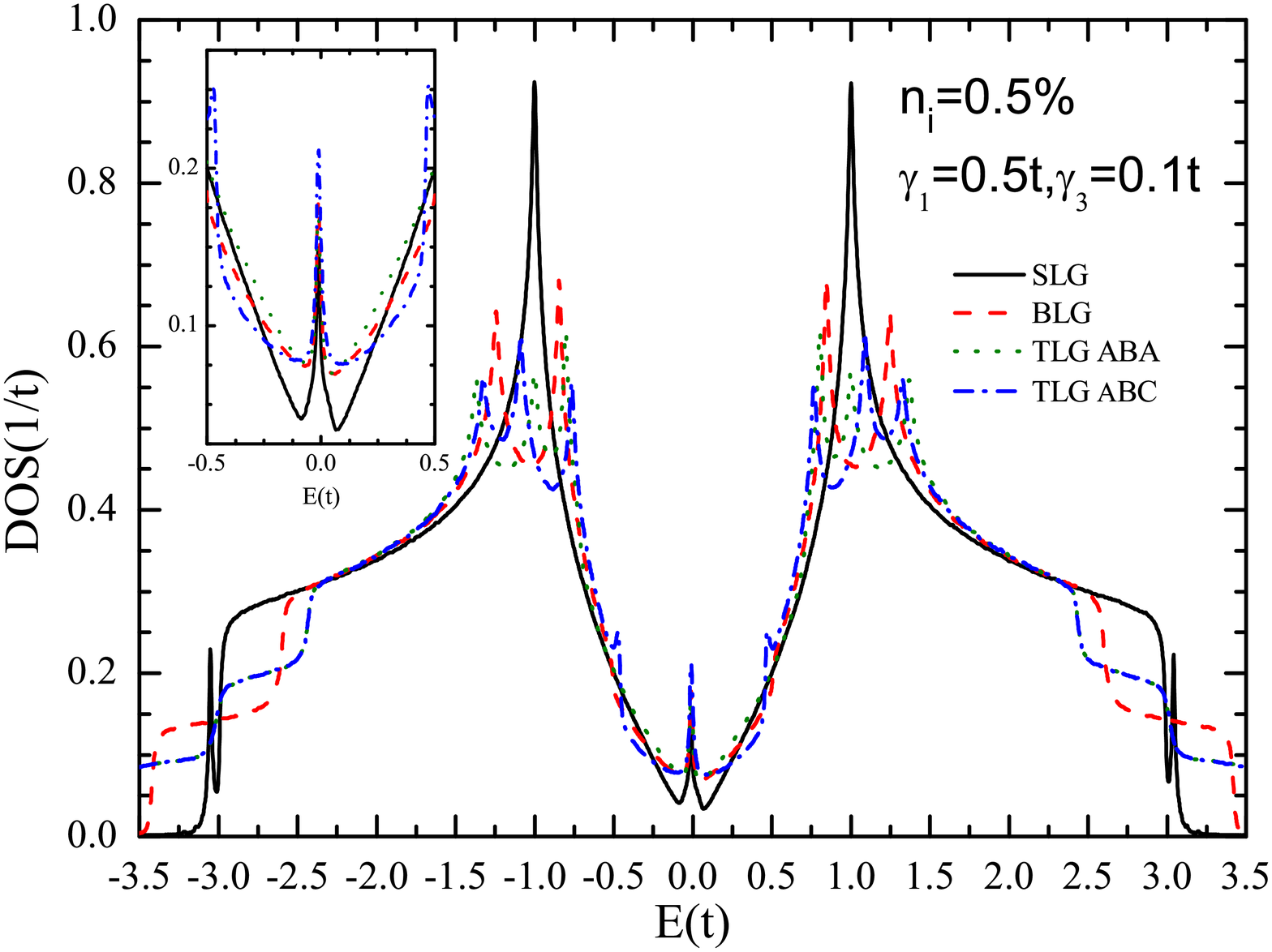}
\includegraphics[width=7.5cm]{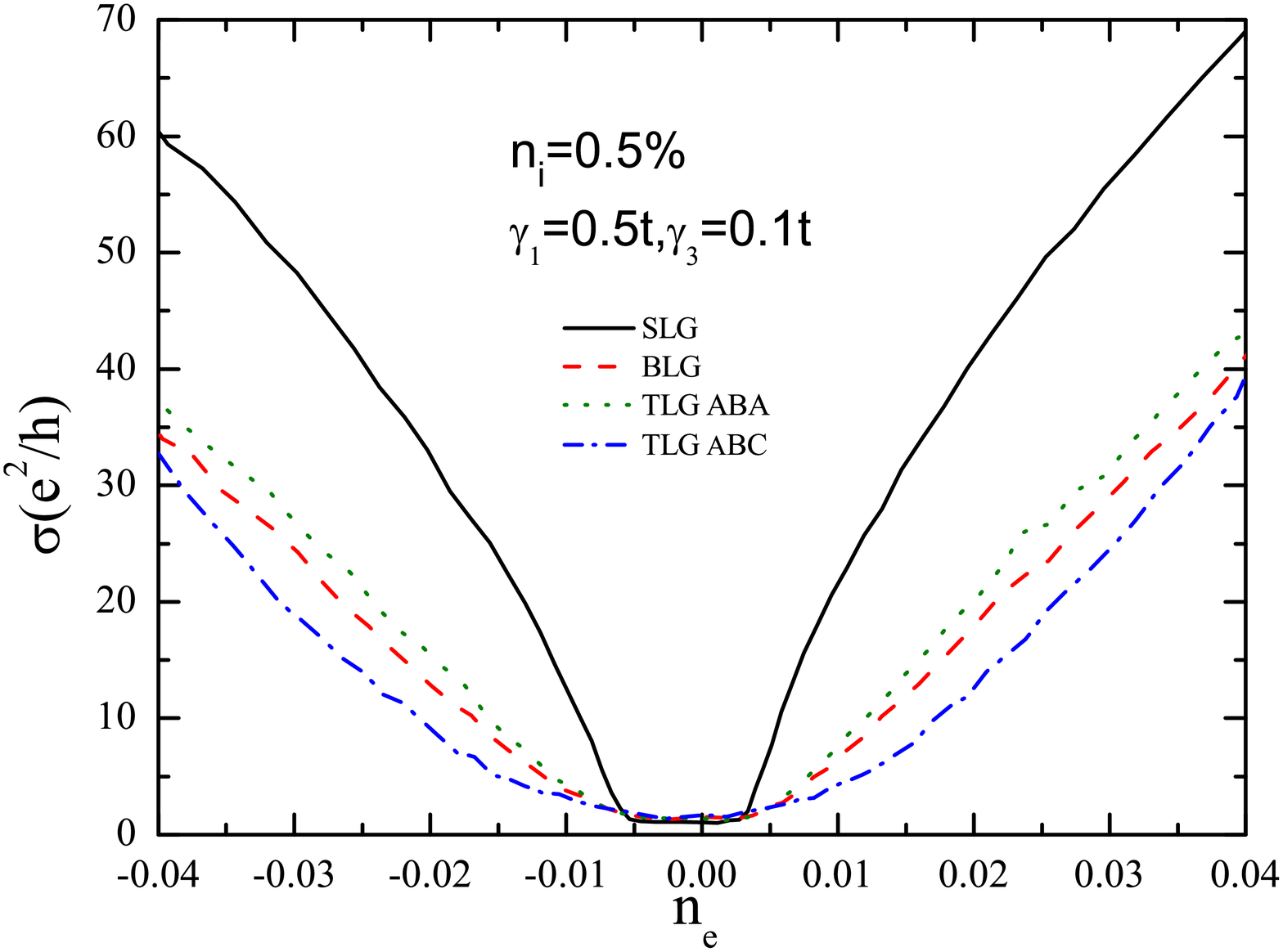}
}
\end{center}
\caption{(Colour online) Comparison of the DOS and conductivity of the SLG,
BLG, TLG and QLG with the same concentration of resonant impurities ($%
n_{i}=0.5\%$, $\protect\varepsilon _{d}=-t/16,$ $V=2t$). $\protect\gamma %
_{1}=\protect\gamma _{3}=0.1t$ in top panels and $\protect\gamma _{1}=0.5t,%
\protect\gamma _{3}=0.1t$ in bottom panels. SLG contains $6400\times 6400\,$%
carbon atoms, each layer in BLG, TLG and QLG contains $4096\times 4096$, $%
3200\times 3200$ and $2400\times 2400$ carbon atoms, respectively.}
\label{ximpslbltri}
\end{figure*}

\begin{figure}[t]
\begin{center}
\includegraphics[width=7.5cm]{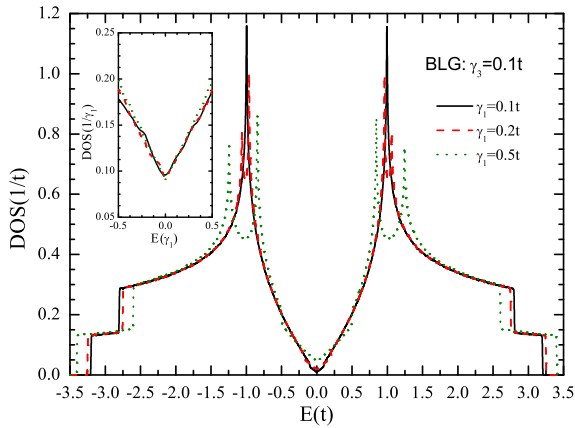}
\end{center}
\caption{(Colour online) Comparison of the DOS of bilayer graphene with
different interlayer interactions: $\protect\gamma _{1}=0.1t$, $0.2t$, and $%
0.5t$ ($\protect\gamma _{3}$ is fixed as $0.1t$). Inner panel: normalized
DOS and energy in units of $1/\protect\gamma _{1}$ and $\protect\gamma _{1}$%
. Each layer in BLG contains $4096\times 4096$ carbon atoms.}
\label{dosblgemma1}
\end{figure}

\begin{figure*}[t]
\begin{center}
\mbox{
\includegraphics[width=7.5cm]{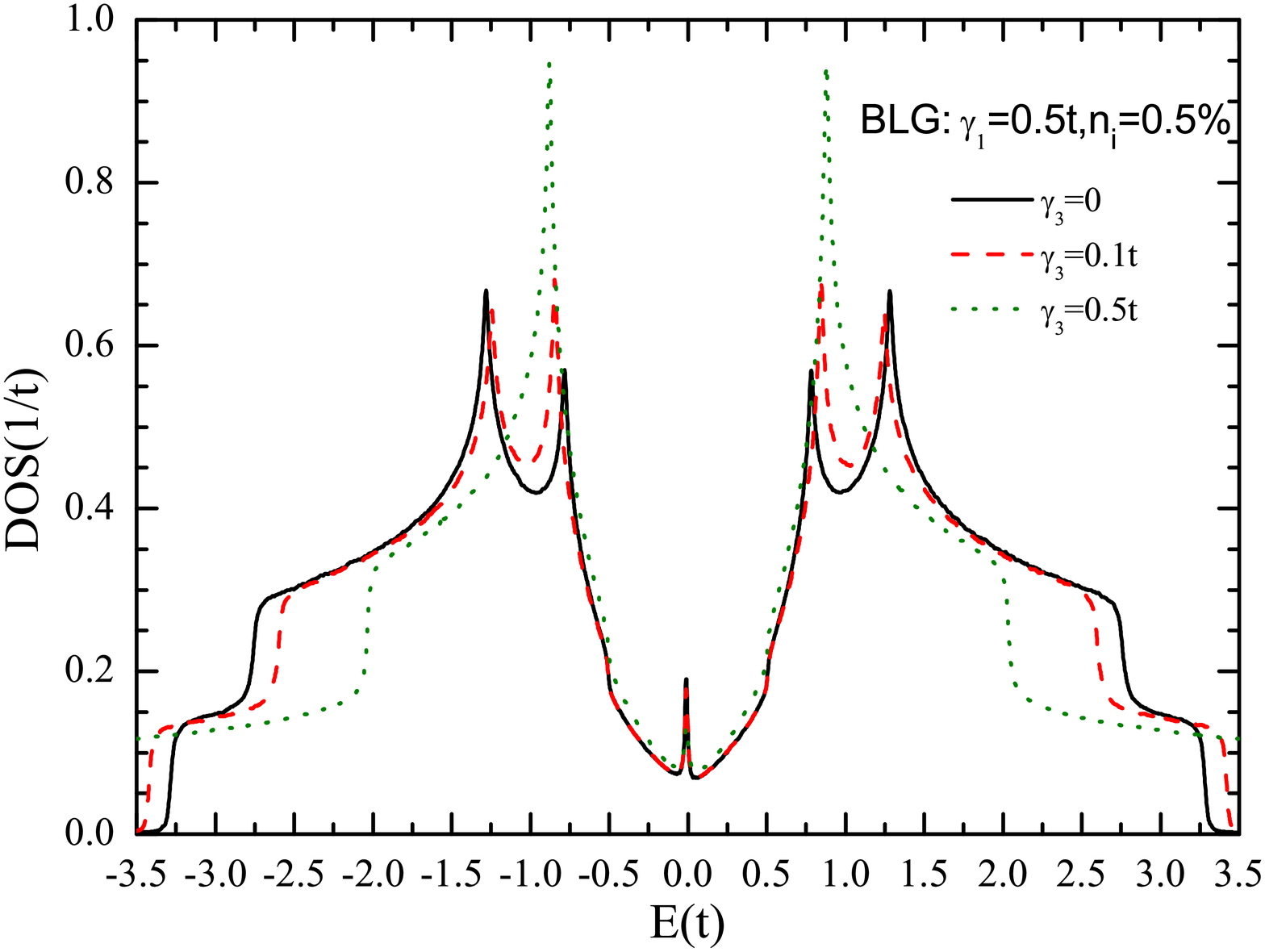}
\includegraphics[width=7.5cm]{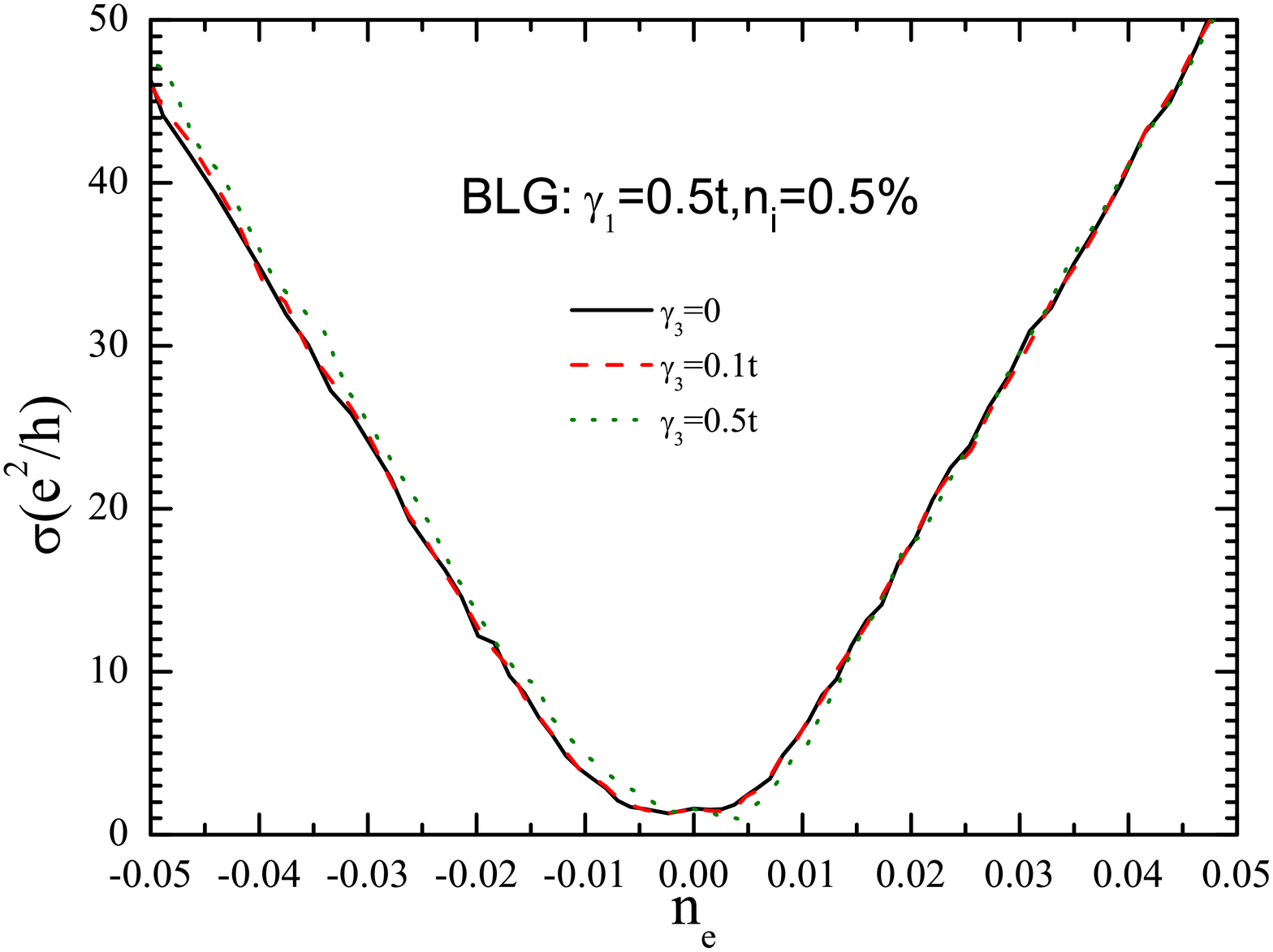}
}
\end{center}
\caption{(Colour online) DOS and conductivity of bilayer graphene with
resonant impurities ($\protect\varepsilon _{d}=-t/16,$ $V=2t$) added on both
layers. The interlayer parameter $\protect\gamma _{1}$ is fixed as $0.5t$,
and $\protect\gamma _{3}=0$ (black line), $0.1t$ (red dashed line) and $0.5t$
(green dot line). Each layer contains $4096\times 4096$ carbon atoms. }
\label{ximpgemma1}
\end{figure*}

\section{Resonant impurities}

Resonant impurities are introduced in reality by the formation of a chemical
band between a carbon atom from graphene sheet and a carbon atom from an
adsorbed organic molecule (CH$_{3}$, C$_{2}$H$_{5}$, CH$_{2}$OH), as well as
H atoms \cite{Wehling2010}; vacancies are another option but in natural
graphene their concentration seems to be small. The adsorbates are described
by the Hamiltonian $H_{imp}$ in Eq.~(\ref{Hamiltonian}). From \textit{ab
initio} density functional theory (DFT) calculations~\cite{Wehling2010}, it
follows that the band parameters for various organic groups (and for
hydrogen atoms) are almost the same: $V\approx 2t$ and $\epsilon _{d}\approx
-t/16$. The hybridization strength $V$ being a factor $2$ larger than $t$ is
in accordance with the hybridization for hydrogen adatoms from Ref. %
\onlinecite{Robinson2008}, but the on-site energies $\epsilon _{d}$ are
significantly smaller than the value $\epsilon _{d}$ $=1.7eV$ used for H in
Ref. \onlinecite{Robinson2008} which makes our results for the transport
properties in SLG qualitatively different \cite{Wehling2010,Yuan2010}. The
adoption of these band parameters successfully explained the resonant
scattering in SLG \cite{Wehling2010,Yuan2010} and we continue to use them in
the modeling of BLG and TLG.

In Refs. \onlinecite{Wehling2010,Yuan2010}, we used the algorithm presented
in the previous section to calculate the dc conductivity of SLG with
resonant impurities or vacancies. We found that there is plateau of the
order of the minimum conductivity \cite{Katsnelson2006} $4e^{2}/\pi h$ in
the vicinity of the neutrality point, in agreement with theoretical
expectations \cite{Ostrovsky2010}. Beyond the plateau, the conductivity is
inversely proportional to the concentration of the impurities, and
approximately proportional to the carrier concentration $n_{e}$. This is
consistent with the approach based on the Boltzmann equation, which in the
limit of resonant impurities with $V\rightarrow \infty $, yields for the
conductivity~\cite{Katsnelson2007,Katsnelson2008,Robinson2008,Wehling2010}
\begin{equation}
\sigma \approx (2e^{2}/h)\frac{2}{\pi }\frac{n_{e}}{n_{i}}\ln ^{2}\left\vert
\frac{E_{F}}{D}\right\vert ,  \label{eqn:imp_lim}
\end{equation}%
where $n_{e}=E_{F}^{2}/D^{2}$ is the number of charge carriers per carbon
atom, and $D$ is of order of the bandwidth. Note that for the case of the
resonance shifted with respect to the neutrality point the consideration of
Ref. \onlinecite{Katsnelson2007} leads to the dependence
\begin{equation}
\sigma \propto \left( q_{0}\pm k_{F}\ln {k_{F}R}\right) ^{2},
\label{fitting}
\end{equation}%
where $\pm $ corresponds to electron and hole doping, respectively, and $R$
is the effective impurity radius. The Boltzmann approach does not work near
the neutrality point where quantum corrections are dominant \cite%
{Ostrovsky2006, Katsnelson2006, Auslender2007}. In the range of
concentrations, where the Boltzmann approach is applicable, our numerical
results of the conductivity of SLG as a function of energy fits very well to
the dependence given by Eq. (\ref{fitting}) \cite{Wehling2010,Yuan2010}.

Electron scattering in BLG has been proven to differ essentially from SLG in
Ref. \onlinecite{Katsnelson_bilayer}: For a scattering potential with radius
much smaller than the de Broglie wavelength of electrons, the phase shift of
$s$-wave scattering $\delta _{0}$ tends to a constant as $k\rightarrow 0$.
Therefore, within the limit of applicability of the Boltzmann equation, the
conductivity of a bilayer should be just linear in $n_{e}$, instead of
sublinear dependence (\ref{fitting}) for SLG. The difference is that in SLG,
due to vanishing DOS at the Dirac point, the scattering disappears at small
wave vectors as $\delta _{0}(k)\propto 1/\ln {kR}$ (with $\ln ^{2}{kR}$ on
the order of 10 for typical amounts of doping) for resonant and as $\delta
_{0}(k)\propto kR$ for the nonresonant impurities. In contrast, in BLG there
are no restrictions on the strength of the scattering and even the unitary
limit ($\delta _{0}=\pi /2$) can be reached at $k=0$.

However, these conclusions are based on the use of an approximate parabolic
spectrum for the bilayer which is valid for the energy interval
\begin{equation}
|E|\ll |\gamma _{1}|.  \label{ineqq1}
\end{equation}%
In the opposite case
\begin{equation}
|E|\gg |\gamma _{1}|  \label{ineqq2}
\end{equation}%
the effects of the interlayer hopping are negligible and one should expect a
behavior of the conductivity similar to that of SLG.

Our first set of numerical calculations of BLG are performed for similar
concentrations of resonant impurities ($n_{i}\in \lbrack 0.1\%,2\%]$) as
those used for SLG in Refs. \onlinecite{Wehling2010,Yuan2010}. The
interlayer hopping parameters are taken as \cite{falko2006} $\gamma
_{1}=\gamma _{3}=0.1t$. As shown in Fig. \ref{ximp}, finite concentrations
of the resonant impurities lead to the formation of a low energy impurity
band (see increased DOS at low energies in Fig. \ref{ximp}). The impurity
band can host two electrons per impurity, and for impurity concentrations in
the range of $[0.1\%,2\%]$, this leads to a plateau-shaped minimum of width $%
2n_{i}$ in the conductivity vs. $n_{e}$ curves around the neutrality point.
As one can see from the DOS in Fig. \ref{ximp}, even for $n_{i}=0.1\%$, the
width of the impurity band around the neutrality point is comparable to the
limits of applicability of the parabolic approximation for the spectrum (\ref%
{ineqq1}), therefore for the concentrations of the impurities presented in
Fig. \ref{ximp} one cannot use the theory \cite{Katsnelson_bilayer}. For
small electron concentrations we are beyond the limit of the Boltzmann
theory at all, and for the larger electron concentration we are, rather, in
the regime (\ref{ineqq2}) so one could expect a sublinear behavior similar
to that in SLG. Indeed, the conductivity of BLG as a function of charge
density $n_{e}$ follows almost exactly the same dependence as for the SLG
(see the direct comparisons of conductivities in Fig. \ref{ximp}). That is,
the density-dependence of conductivity in BLG is not linear but sublinear (%
\ref{fitting}) as in SLG. Actually, as shown in Fig. \ref{ximpslbltri}, the
sublinear dependence is quite general for multilayer graphene, i.e., it is
also true for trilayer and quartic-layer graphene with the same
concentration of resonant impurities, independent on the stacking sequence,
which is, of course, not surprising assuming that the condition (\ref{ineqq2}%
) holds. Here for the trilayer (quartic-layer) we consider two types of
stacking sequence: ABA (ABAB) and ABC (ABCD). This general property of the
conductivity can be easily understood by comparison of their DOS in Fig. \ref%
{ximpslbltri}. The DOS of single-layer, bilayer, trilayer and quartic-layer
graphene are exactly the same except near the edge of the spectrum,
indicating the similar band structure, independent on the number of layers
and stacking sequence. In fact, since the couplings between the carbon atoms
and organic admolecules are twenty times larger than the interlayer coupling
($V=20\gamma _{1}$) in our model, the unique bonds generated by the relevant
weaker interlayer interactions are more easily to be destroyed by the
impurity bonds generated by the much stronger adsorbed resonant impurities.

\begin{figure*}[t]
\begin{center}
\mbox{
\includegraphics[width=7.5cm]{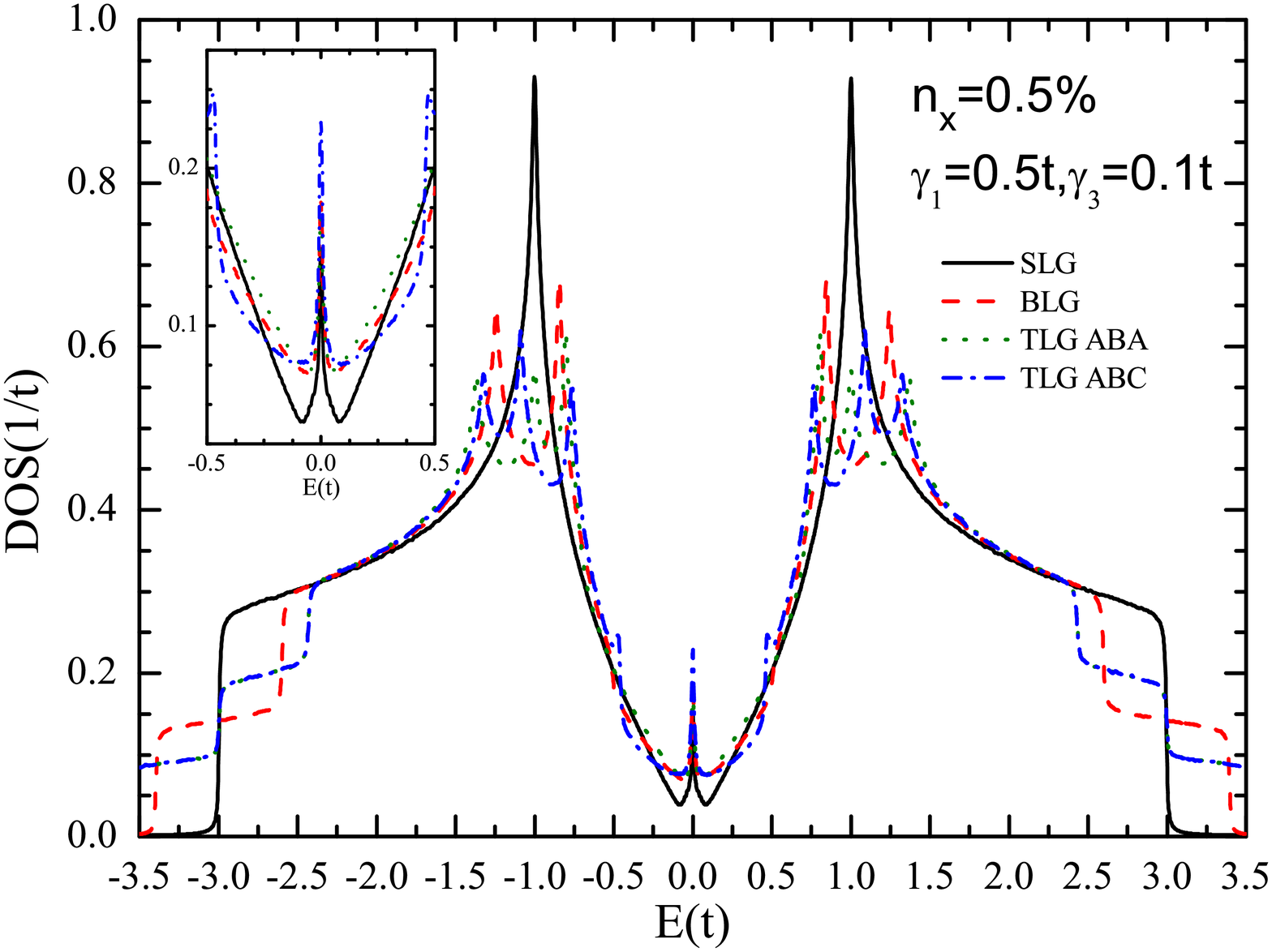}
\includegraphics[width=7.5cm]{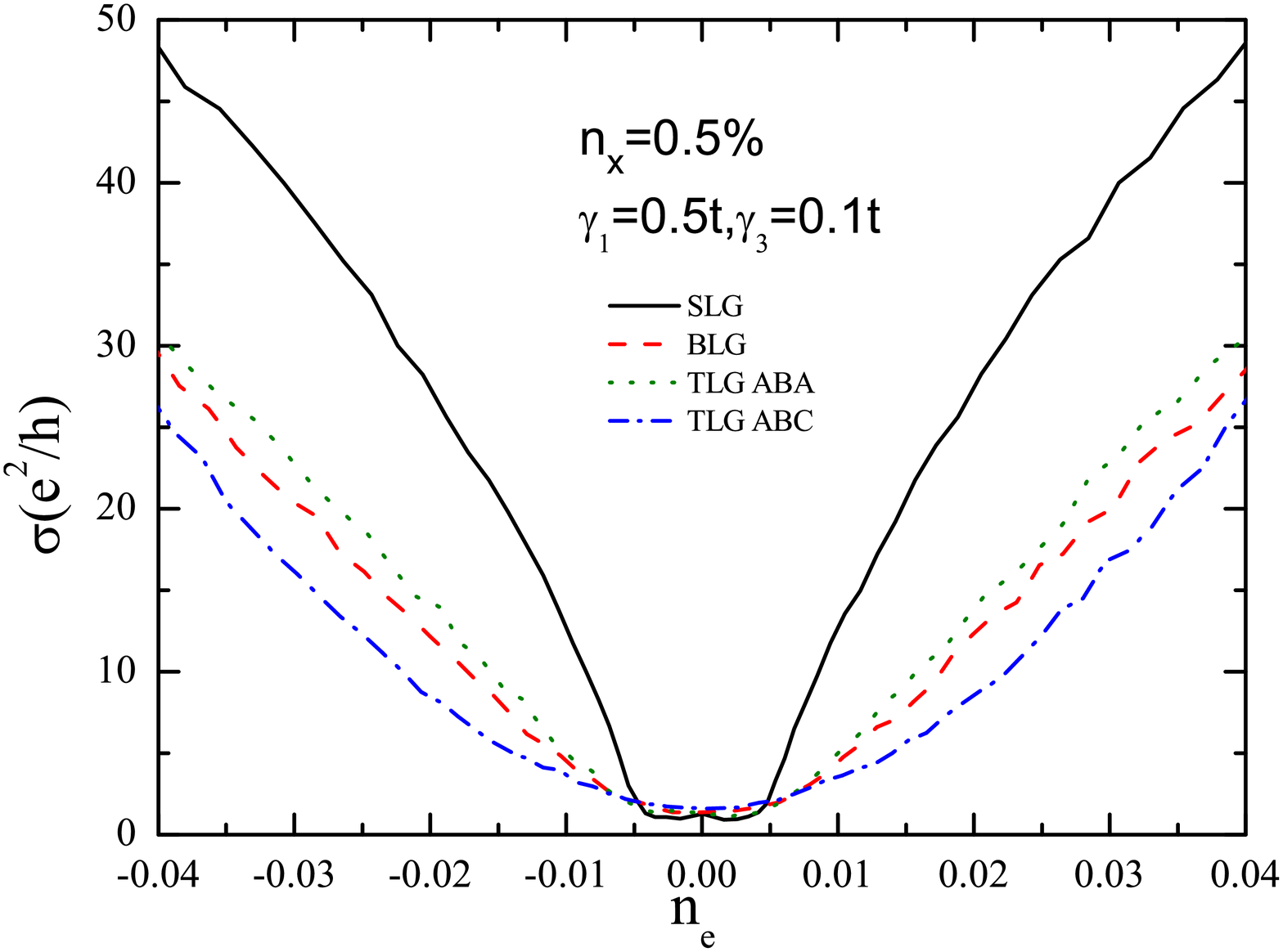}
}
\end{center}
\caption{(Colour online) Comparison of the DOS and conductivity of the SLG,
BLG and TLG with the same concentration of vacancies ($n_{x}=0.5\%$). The
parameters of the interlayer coupling are $\protect\gamma _{1}=0.5t$ and $%
\protect\gamma _{3}=0.1t$. SLG contains $6400\times 6400\,$carbon atoms,
each layer in BLG and TLG contains $4096\times 4096$ and $3200\times 3200$
carbon atoms (sites), respectively.}
\label{dcvacancy}
\end{figure*}

In order to check the symmetry of the presence of the impurities, we limit
the adsorption of organic admolecules to one layer of BLG ($n_{1i}$, case
II). To compare the results of the adsorption on both sides ($n_{i}$, case
I), we fix the total number of resonant impurities and therefore the
concentration on one layer (case II) is doubled ($n_{1i}=N_{imp}/N_{carbon%
\_in\_one\_layer}=2n_{i}$). We find (see last panel in Fig. \ref{ximp}) that
for the low concentration ($n_{i}\leq 0.5\%$), the electron-density
dependence of the conductivity in BLG follows the same law in both cases; at
high concentration ($n_{i}\geq 1\%$), the conductivity in case II is larger
than in case I. This is because in case II the difference of mobility of
electron in the two layers, with or without impurities, is larger than in
the case of large concentrations of adsorbed admolecules.

Next we consider the region of parameters which can be described by the
Boltzmann equation plus parabolic spectrum \cite{Katsnelson_bilayer}. In
BLG, the approximations of massive valence and conduction bands with zero
gap: $E\left( k\right) =\pm \hbar ^{2}k^{2}/2m^{\ast }$, where the effective
mass is given by $m^{\ast }=\gamma _{1}/2v_{F}^{2}$, are only true in the
low-energy dispersion close to the neutrality point. There are two ways to
place the impurity bands within the region of low-energy dispersion (\ref%
{ineqq1}): decreasing the concentration $n_{i}$ of impurities or expanding
the quadratic band by increasing $\gamma _{1}$. Smaller concentration of
impurities leads to less random states for the averaging in the Kubo formula
of Eq. (\ref{conducappr}), which means that it is numerically more expensive
because we need to extend the sample size to keep the same accuracy.
Therefore increasing $\gamma _{1}$ is computationally more convenient from
the point of view of CPU time and physical memory; one can assume that
physical results should be the same: it is only the ratio $n_{i}/\gamma _{1}$
which is important.

In Fig. \ref{dosblgemma1}, we compare numerical results of DOS of BLG with
different band parameter $\gamma _{1}$: $0.1t,0.2t$ and $0.5t$ ($\gamma _{3}$
is fixed as $0.1t$). One can see that the width of the parabolic band with
the energy-independent local density of states proportional to $\gamma _{1}$%
, and the normalized energy (in units of $\gamma _{1}$) dependencies of DOS
(in the units of $1/\gamma _{1}$) within the parabolic band are consistent
for different $\gamma _{1}$ (see the inner panel of Fig. \ref{dosblgemma1}).
Therefore we can simply use $\gamma _{1}$ with the value of $0.5t$ instead
of $0.1t$ to extend the width of the parabolic band approximation without
changing the structure of the spectrum. The numerical results form a system
with $m$ times larger of $\gamma _{1}$, are qualitatively comparable to
those for a system of $1/m$ times smaller concentration $n_{i}$ of
impurities.

The numerical results for the DOS and conductivities of BLG and TLG in the
presence of resonant impurity with larger interlayer interactions ($\gamma
_{1}=0.5t,$) are shown in Fig. \ref{ximpslbltri}. We see that for an
impurity concentration of $n_{i}=0.5\%$, the impurity band is located around
the neutral point and far from the edge of the quadratic band ($\left\vert
E\right\vert <0.5t$). In the region of the impurity band ($|n_{e}|\leq
n_{i}=0.5\%$), there is a plateau in the order of $2e^{2}/h$ (per layer) in
BLG, as well as in TLG. This values is slightly larger than the minimum
conductivity $4e^{2}/\pi h$ of SLG. It is worthwhile to note that an
explanation of the origin of plateau around the neutrality point is beyond
the applicability of Boltzmann equation, just as in the case of SLG \cite%
{Wehling2010,Yuan2010}. Analyzing experimental data of the plateau width
(similar to the analysis for N$_{2}$O$_{4}$ acceptor states in Ref. %
\onlinecite{Wehling2008}) can therefore yield an independent estimate of the
impurity concentration, both in single-layer and multilayer graphene. Within
the parabolic band but beyond the impurity band, the conductivities in BLG
and ABA-stacked TLG exhibit very well the linear dependence on the charge
density $n_{e}$. The ABC-stacked TLG is different from the others because of
its unique band structures with a cubic touching of the bands \cite{r3} (see
the difference of DOS in Fig. \ref{ximpslbltri}).

Finally we check the role of $\gamma _{3}$ in the conductivity of BLG.
Theoretically, the influence of $\gamma _{3}$ to the band structure is
negligible, and so it is for the conductivity. This is confirmed by our
numerical results in Fig. \ref{ximpgemma1}. For the fixed concentration of
impurities $n_{i}=0.5\%$ with $\gamma _{1}=0.5t$, the values of the
conductivity corresponding to the same electron concentration $n_{e}$ are
quite close for $\gamma _{3}=0$, $0.1t$, and $0.5t$.

\begin{figure*}[t]
\begin{center}
\mbox{
\includegraphics[width=7.5cm]{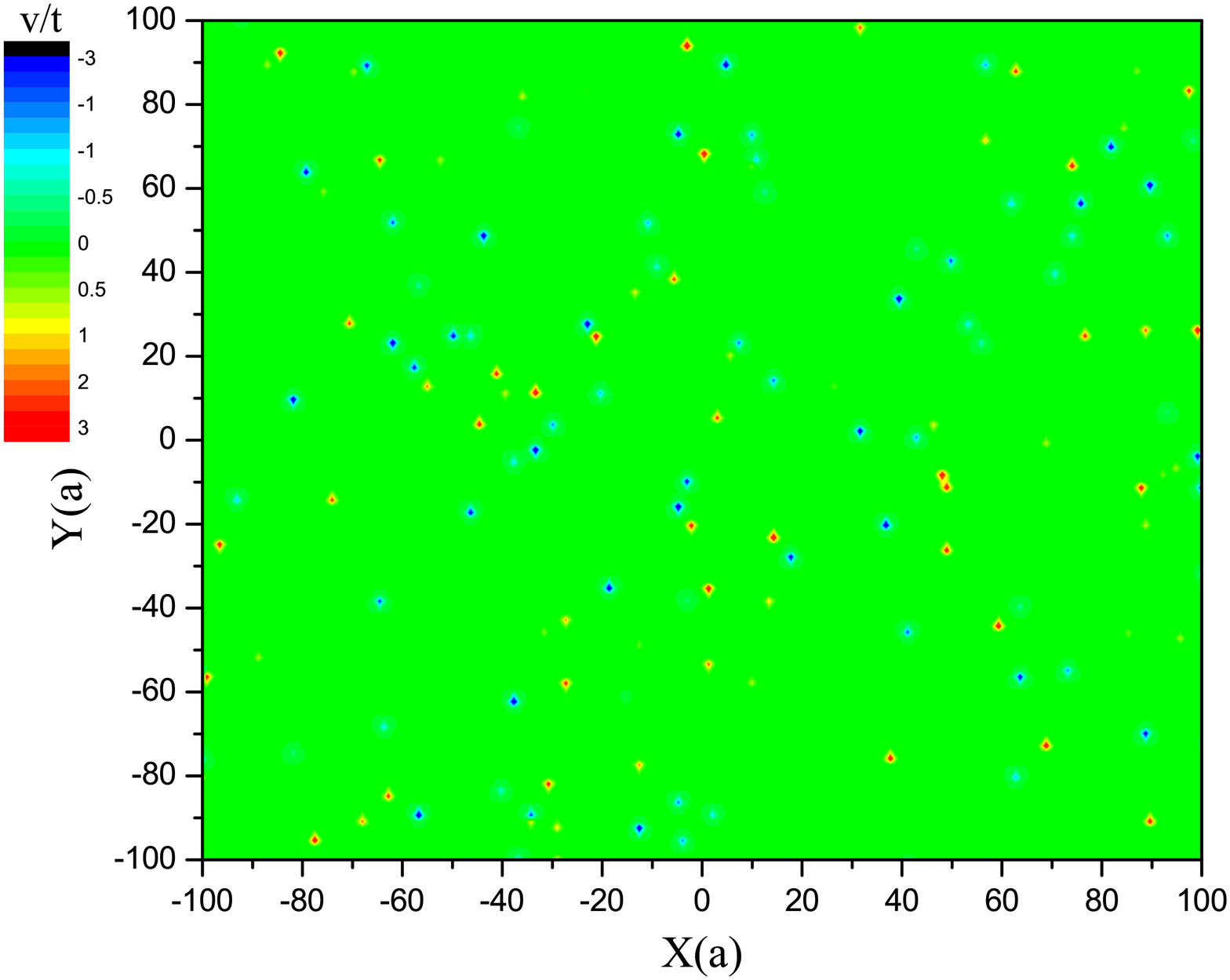}
\includegraphics[width=7.5cm]{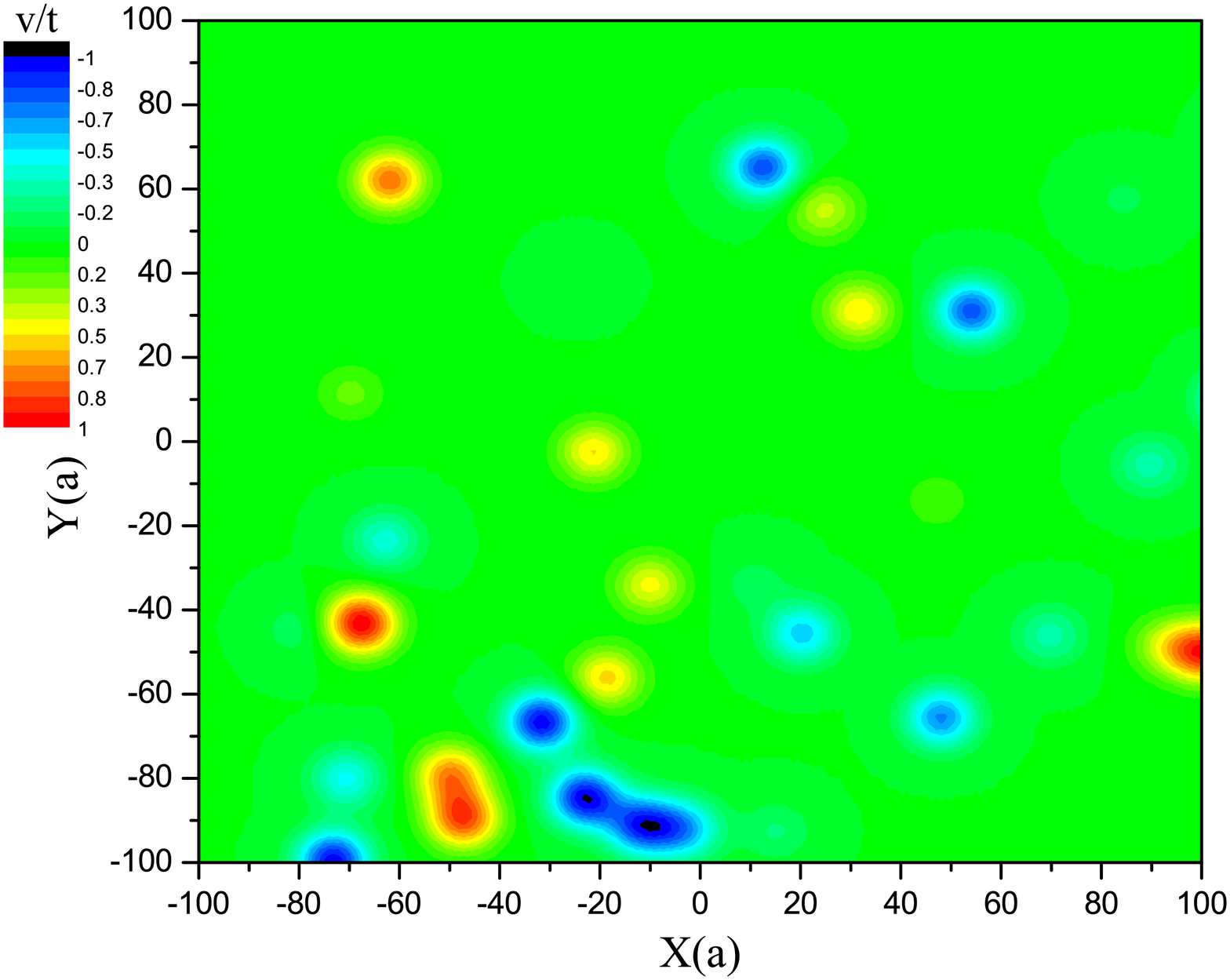}
}
\end{center}
\caption{(Colour online) Contour plot of the on-site potentials in the
central part of a graphene layer ($4096\times 4096$) with short-range ($%
\Delta =3t,$ $d=0.65a,$ $P_{v}=0.5\%$) or long-range ($\Delta =1t,$ $d=5a,$ $%
P_{v}=0.1\%$) Gaussian potential.}
\label{contourpotential}
\end{figure*}

\section{Vacancies}

A vacancy in a graphene sheet can be regarded as an atom (lattice point)
with an on-site energy $v\rightarrow \infty $ or with its hopping parameters
to other sites being zero. In the numerical simulation, the simplest way to
implement a vacancy is to remove the atom at the vacancy site. Introducing
vacancies in SLG will create a zero energy modes (midgap state) \cite%
{Peres2006,Pereira2006,Pereira2008,Wehling2010,Yuan2010}. The exact
analytical wave function associated with the zero mode induced by a single
vacancy in SLG was obtained in Ref. \onlinecite{Pereira2006}, showing a
quasilocalized character with the amplitude of the wave function decaying as
inverse distance to the vacancy. SLG with a finite concentration of
vacancies was studied numerically in Refs. %
\onlinecite{Peres2006,Pereira2008,Wehling2010,Yuan2010,Jafri2010,Chang2010,Wu2010}%
. It was shown that the number of the midgap states increases with the
concentration of the vacancies \cite%
{Peres2006,Pereira2008,Wehling2010,Yuan2010}, and \textit{quasieigenstates}
are also quasilocalized around the vacancies \cite{Yuan2010}. The inclusion
of vacancies brings an increase of spectral weight to the surrounding of the
Dirac point ($E=0)$ and smears the van Hove singularities \cite%
{Peres2006,Pereira2008,Wehling2010,Yuan2010}. The effect of the vacancies on
the transport properties of SLG is quite similar to that of the adsorbed
organic molecules. The main difference is the position of the impurity band
in the spectrum: its center is located at the neutrality point in the
presence of vacancies, whereas it is biased in the presence of realistic
resonant impurities because of the nonzero on-site potential on the organic
carbon (or hydrogen) atom. The vacancy band contributes to the conductivity
and leads to a plateau of minimum conductivity in the midgap region. The
width of the plateau is $2n_{x}$ ($n_{x}$ is the concentration of the
vacancies) in the conductivity vs. $n_{e}$ curves around the neutrality
point, showing the same dependence ($2n_{i}$) as the case of resonant
impurities \cite{Wehling2010,Yuan2010}. For the range of concentrations
where the Boltzmann approach is applicable, the conductivity of SLG as a
function of energy fits very well to the dependence given by Eq. (\ref%
{fitting}), with $q_{0}=0$ for the vacancies and $q_{0}\neq 0$ for the
resonant impurities \cite{Wehling2010,Yuan2010}.

\begin{figure*}[t]
\begin{center}
\mbox{
\includegraphics[width=7.5cm]{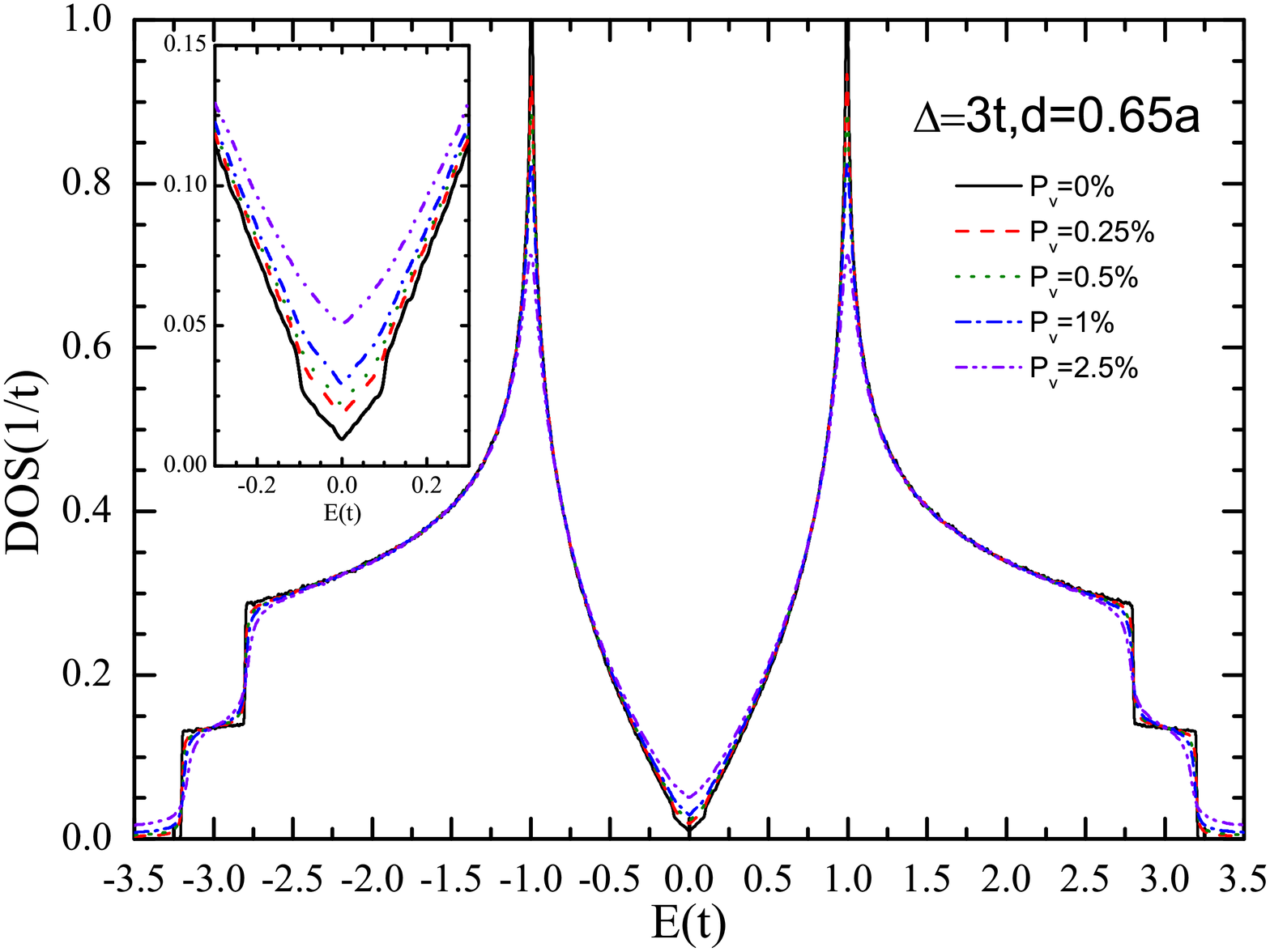}
\includegraphics[width=7.5cm]{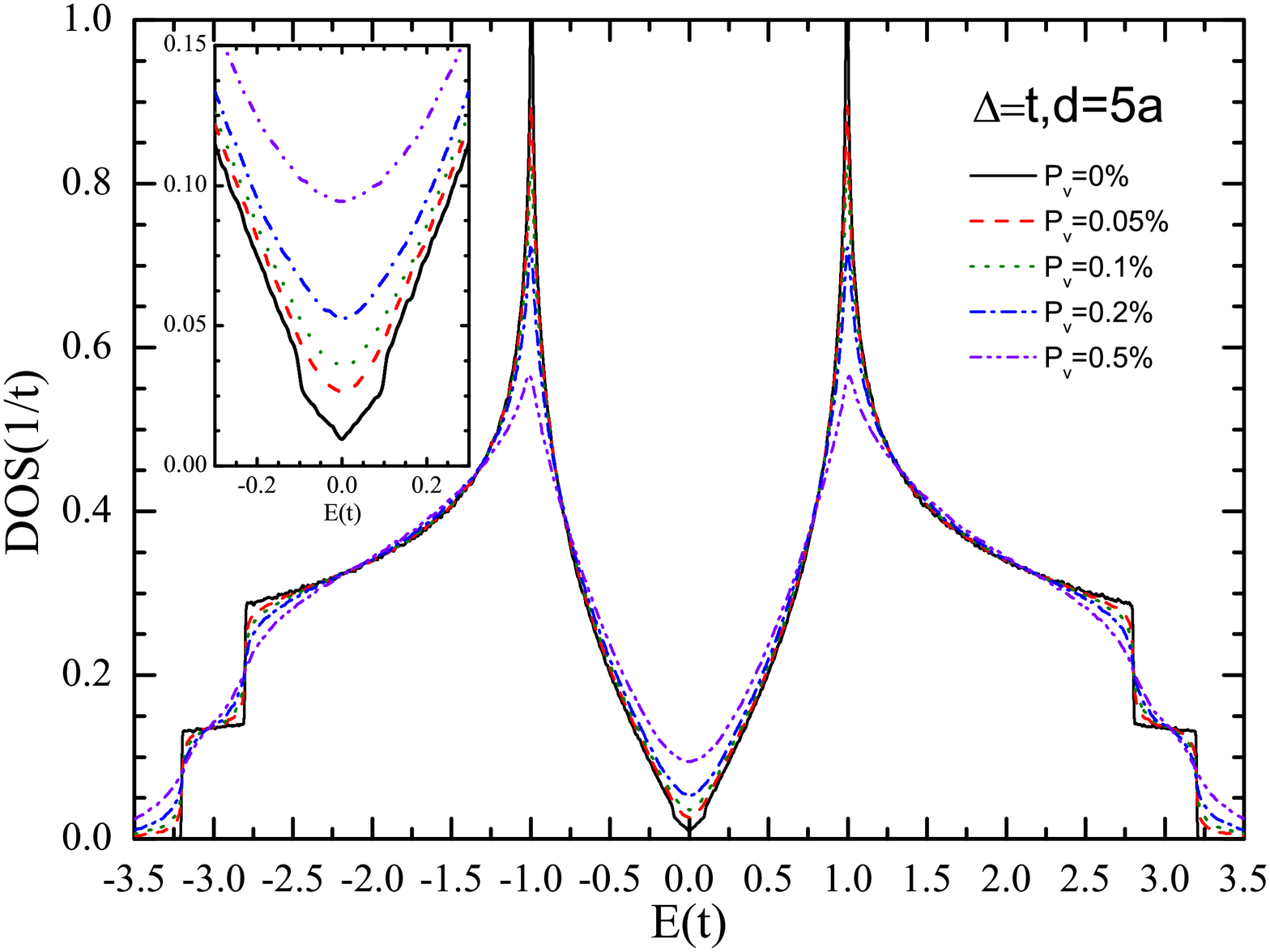}
}
\mbox{
\includegraphics[width=7.5cm]{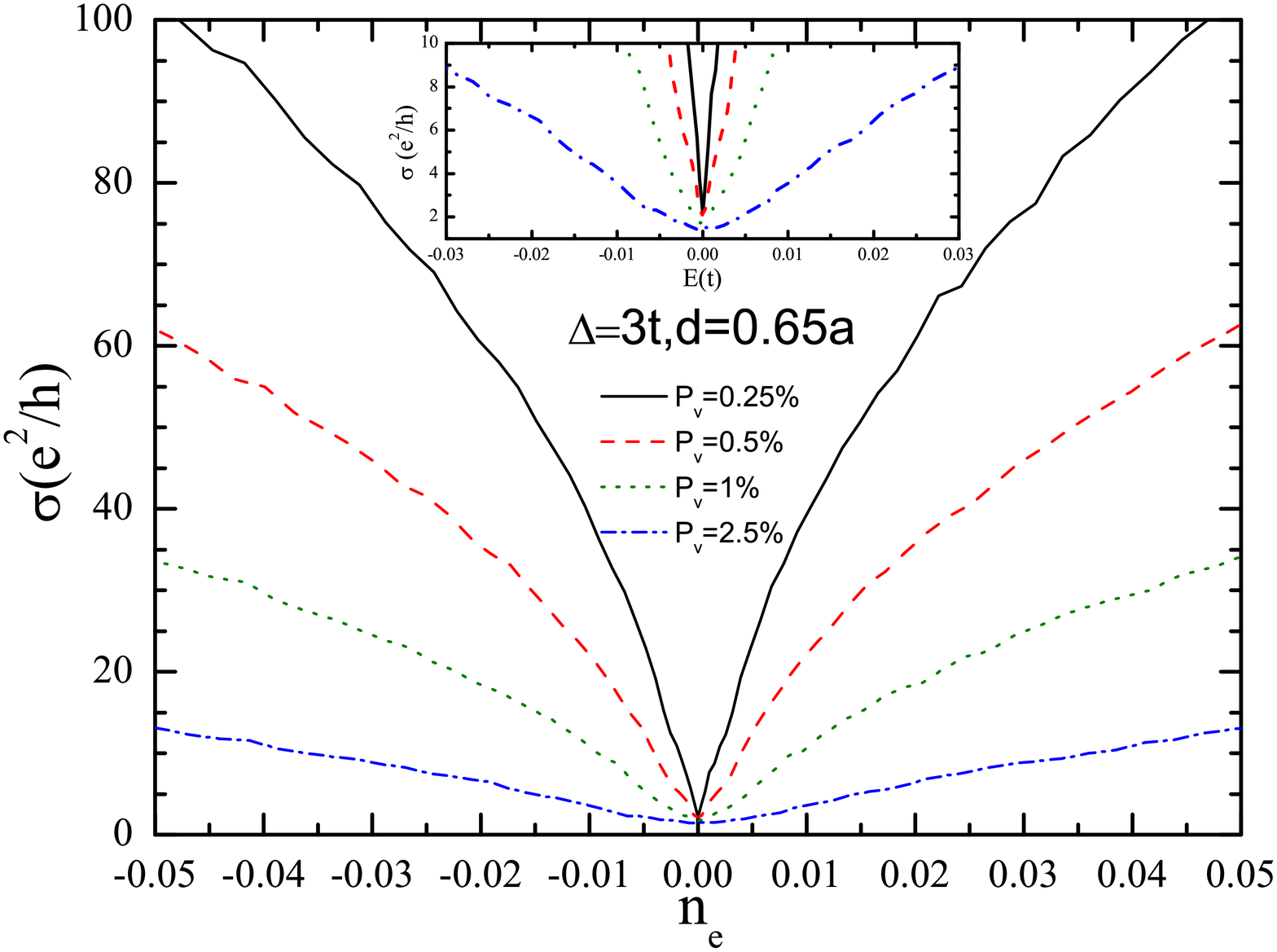}
\includegraphics[width=7.5cm]{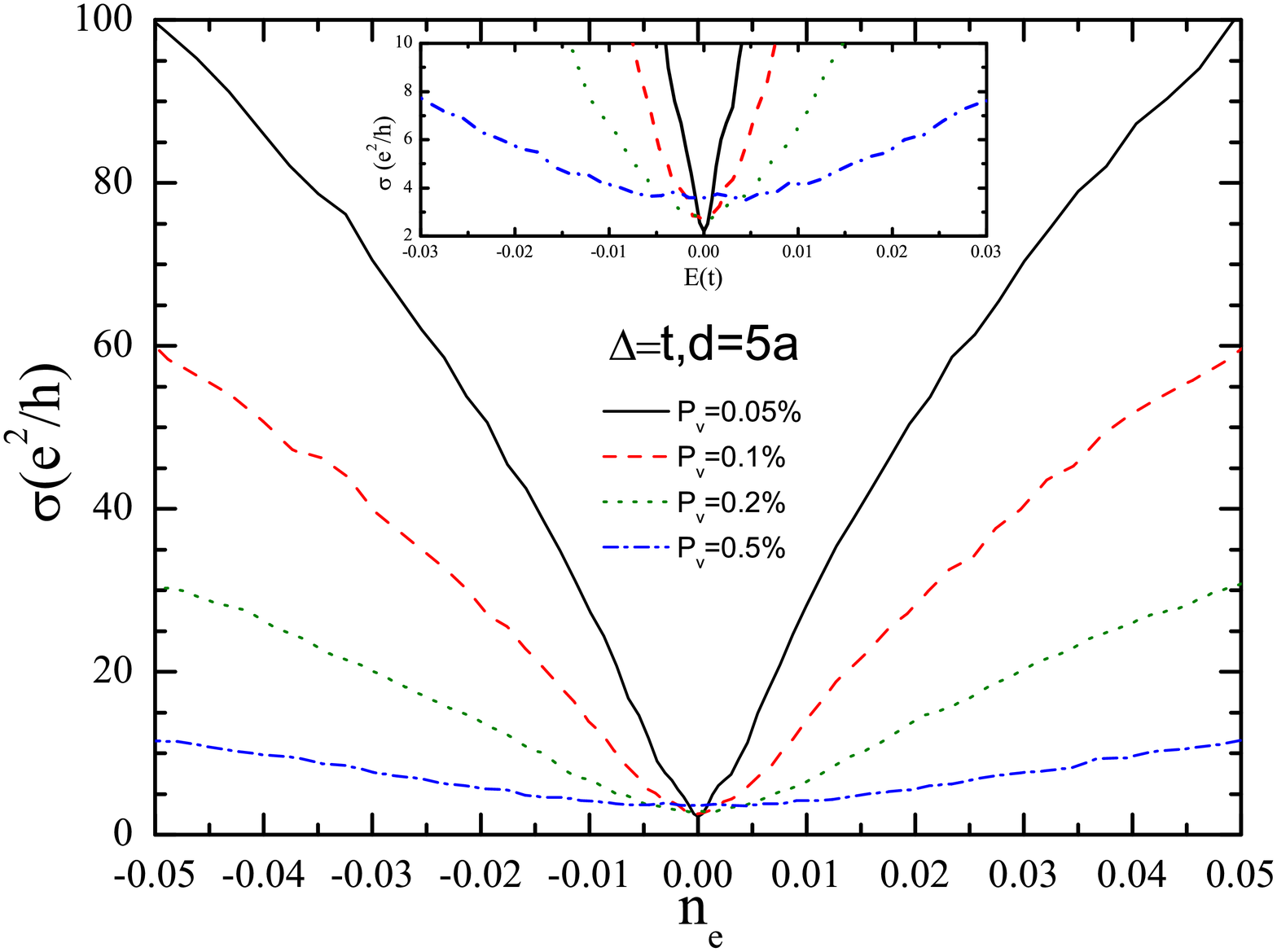}
}
\end{center}
\caption{(Colour online) DOS and conductivity of bilayer graphene ($\protect%
\gamma _{1}=\protect\gamma _{3}=0.1t$) with short-range ($\Delta =3t,$ $%
d=0.65a$) or long-range ($\Delta =1t,$ $d=5a$) Gaussian potential. Each
layer contains $4096\times 4096$ carbon atoms.}
\label{gaussian}
\end{figure*}

\begin{figure*}[t]
\begin{center}
\mbox{
\includegraphics[width=7.5cm]{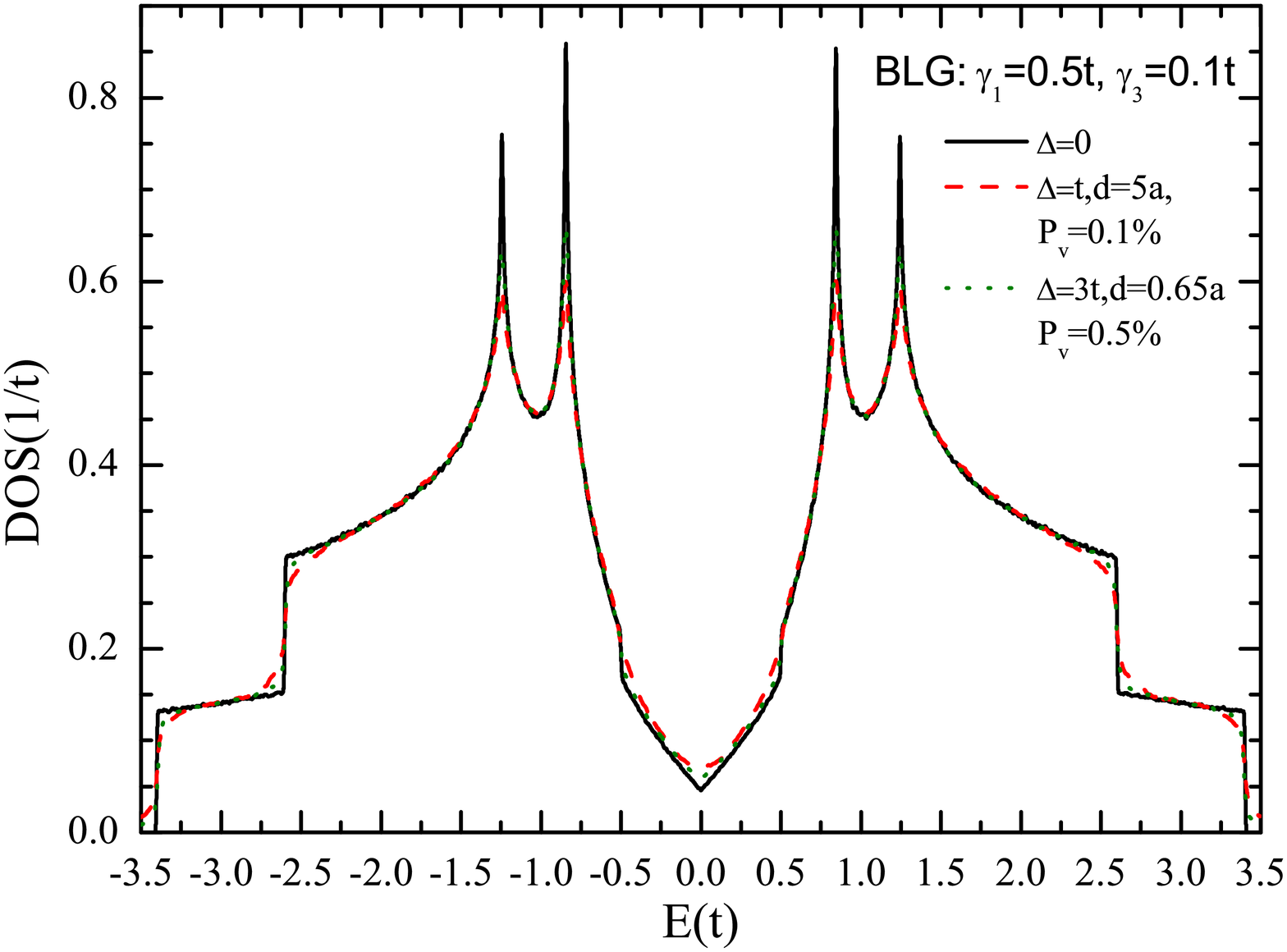}
\includegraphics[width=7.5cm]{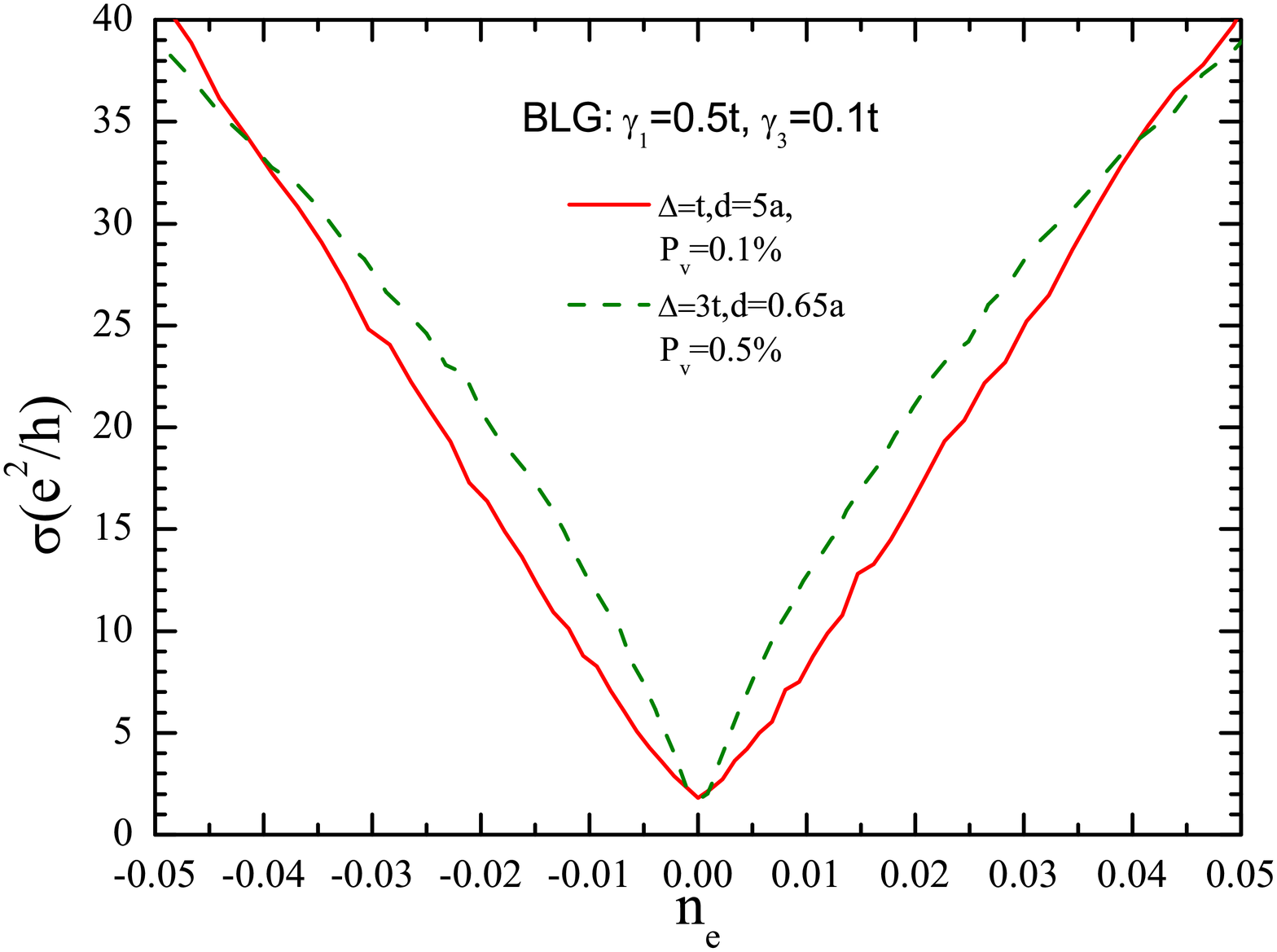}
}
\mbox{
\includegraphics[width=7.5cm]{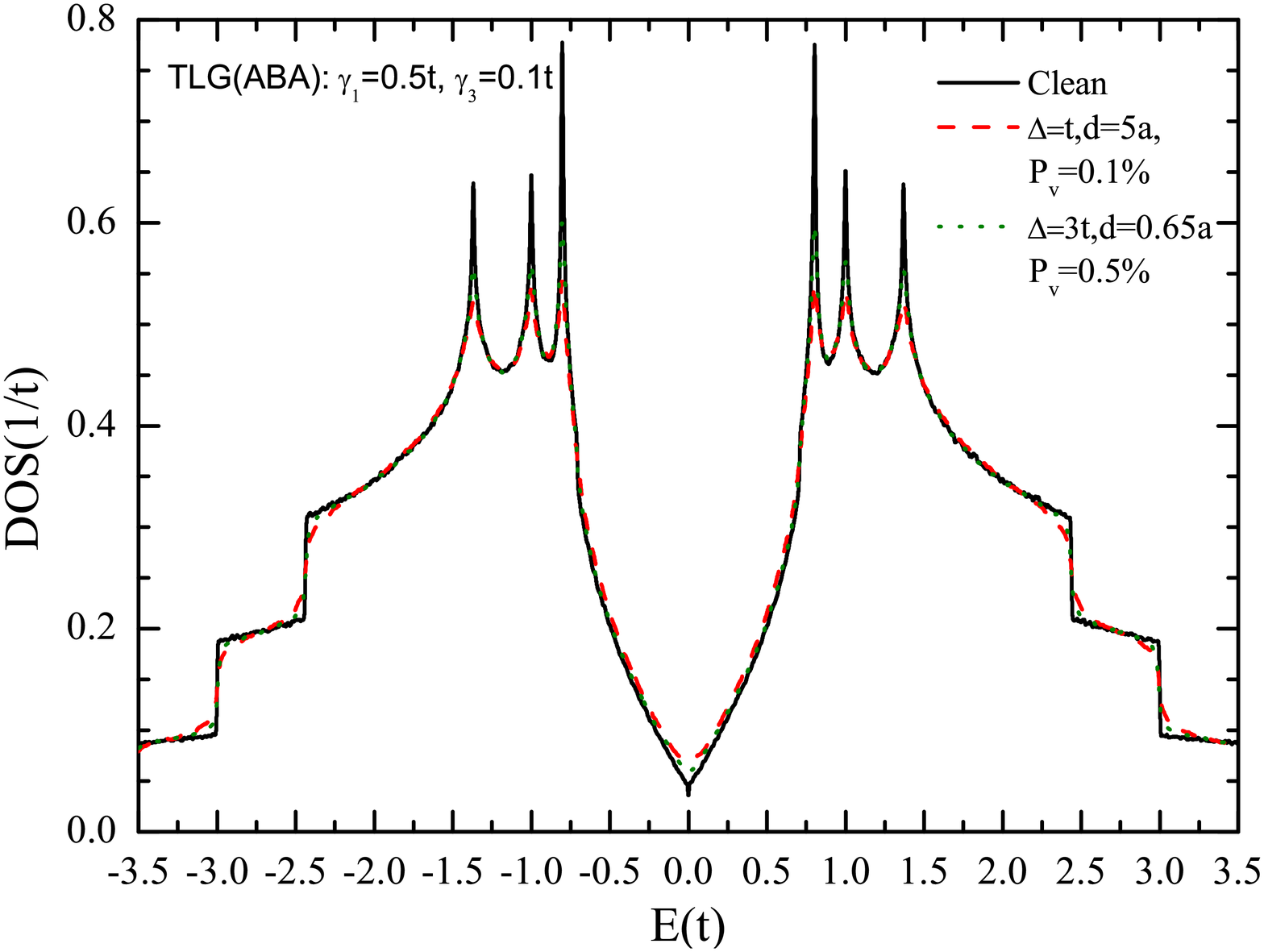}
\includegraphics[width=7.5cm]{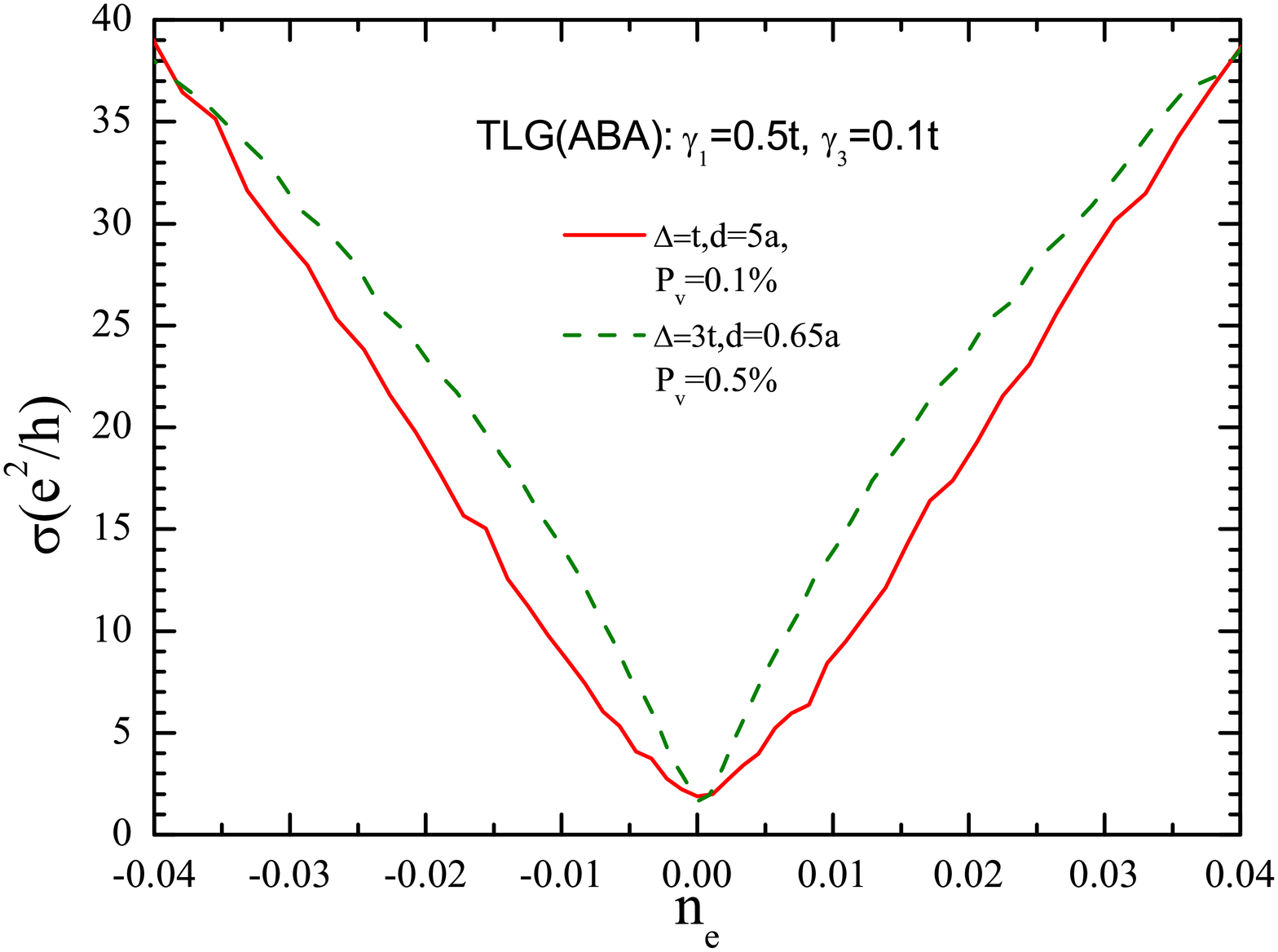}
}
\mbox{
\includegraphics[width=7.5cm]{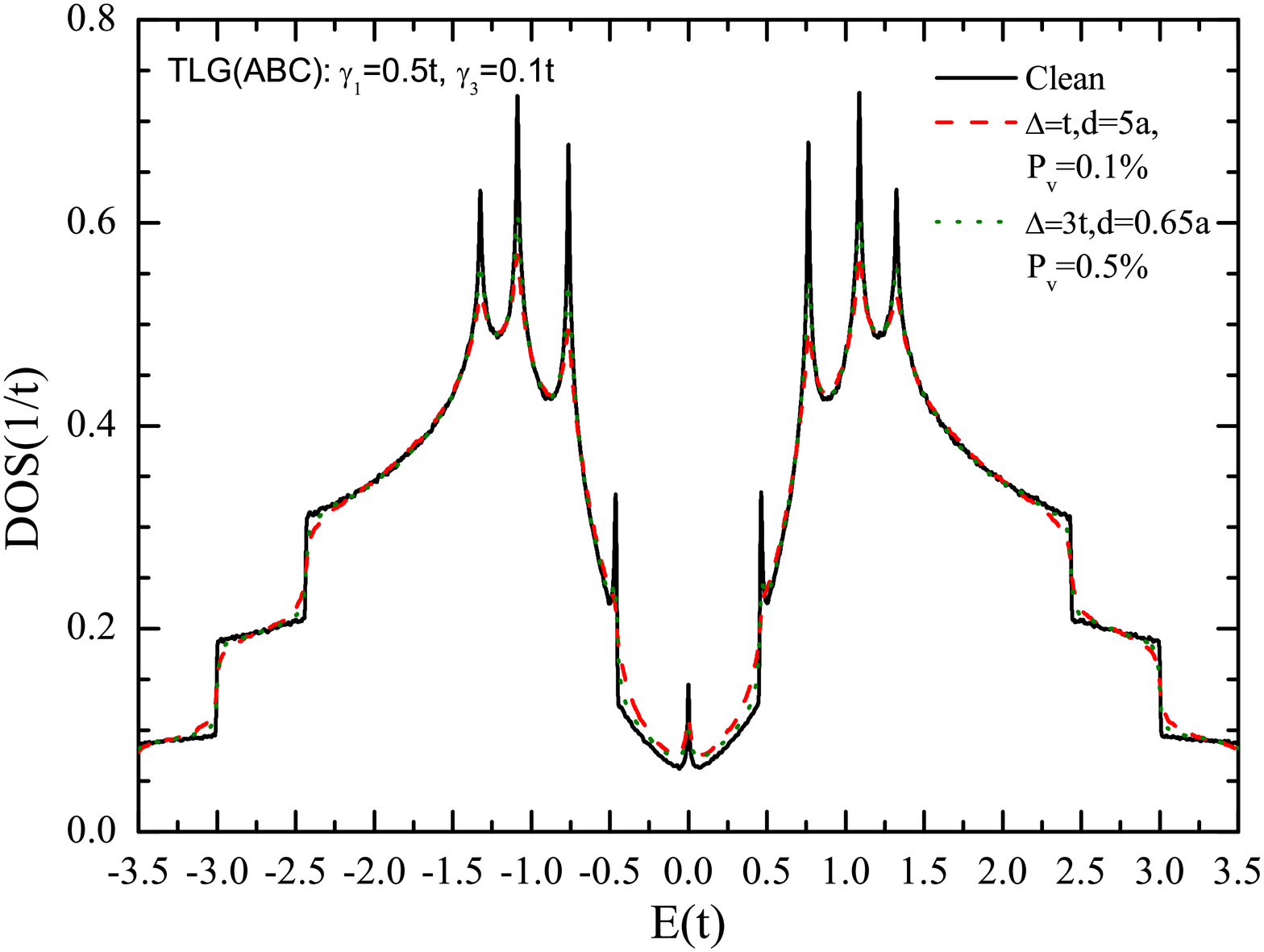}
\includegraphics[width=7.5cm]{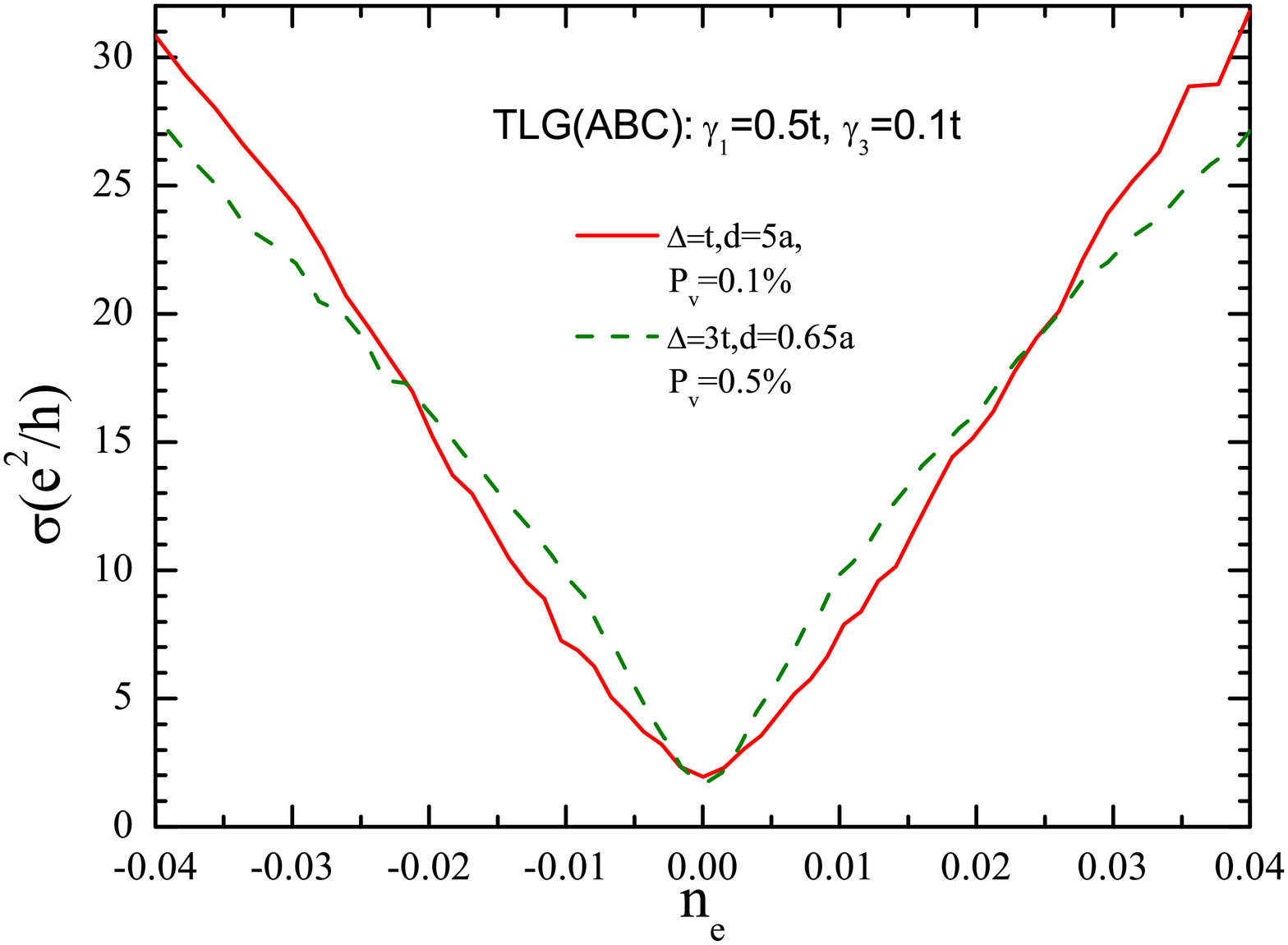}
}
\end{center}
\caption{(Colour online) DOS and conductivity of bilayer graphene ($\protect%
\gamma _{1}=0.5t$, $\protect\gamma _{3}=0.1t$) with long-range ($\Delta =1t,$
$d=5a$) or short-range ($\Delta =3t,$ $d=0.65a$) Gaussian potential. Each
layer contains $4096\times 4096$ carbon atoms.}
\label{gaussian_gemma1}
\end{figure*}

\begin{figure*}[t]
\begin{center}
\mbox{
\includegraphics[width=7.5cm]{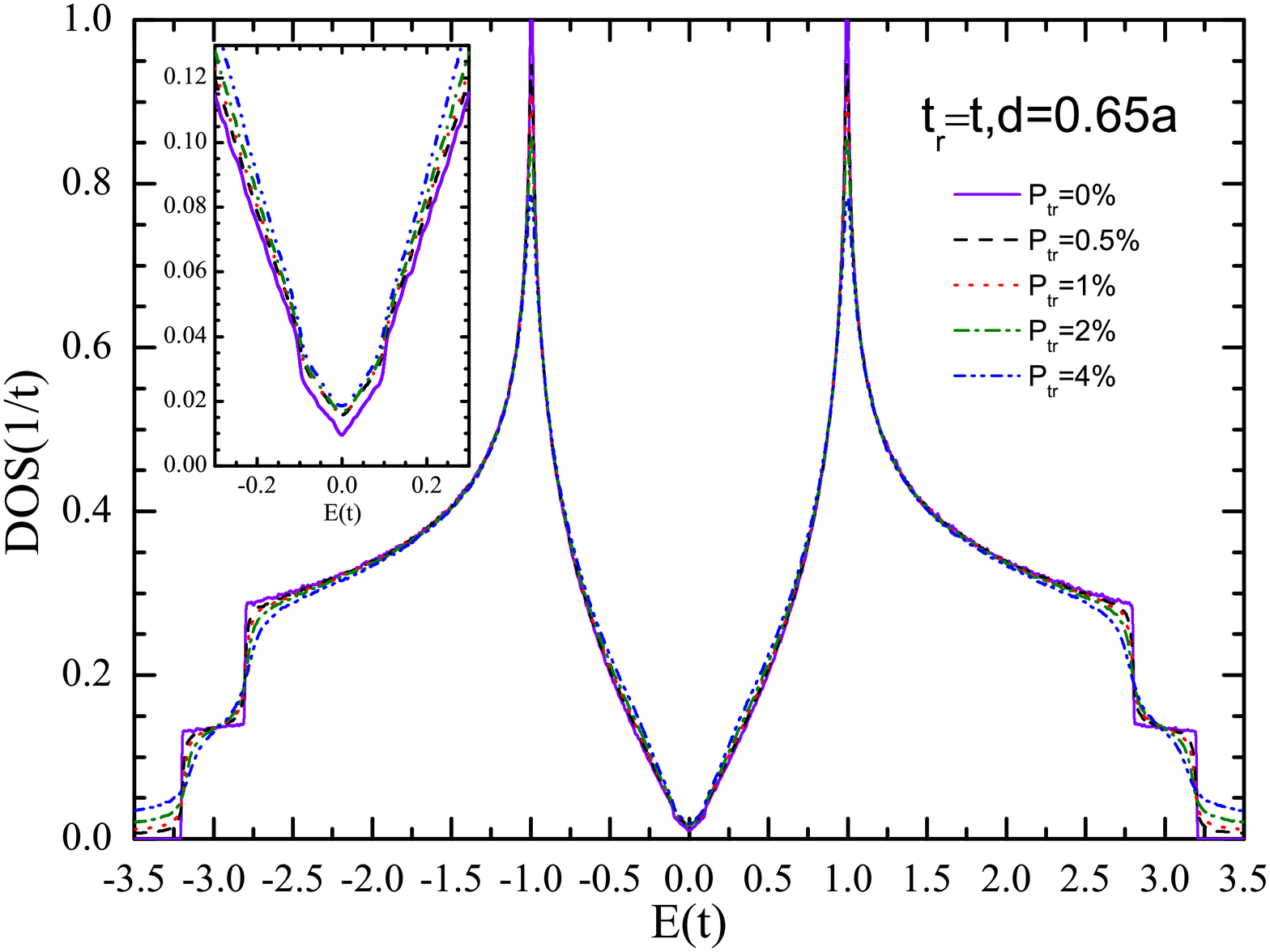}
\includegraphics[width=7.5cm]{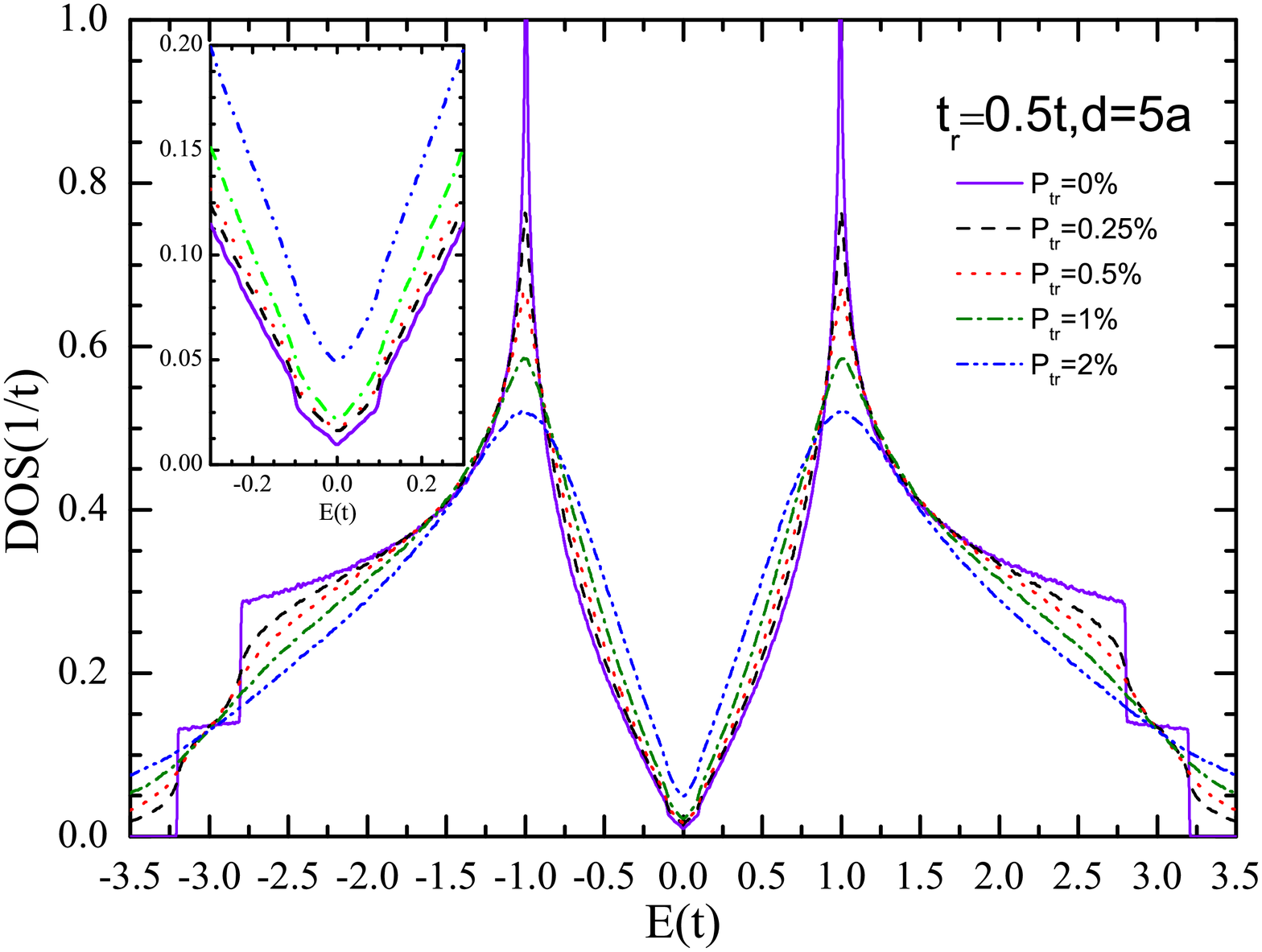}
}
\mbox{
\includegraphics[width=7.5cm]{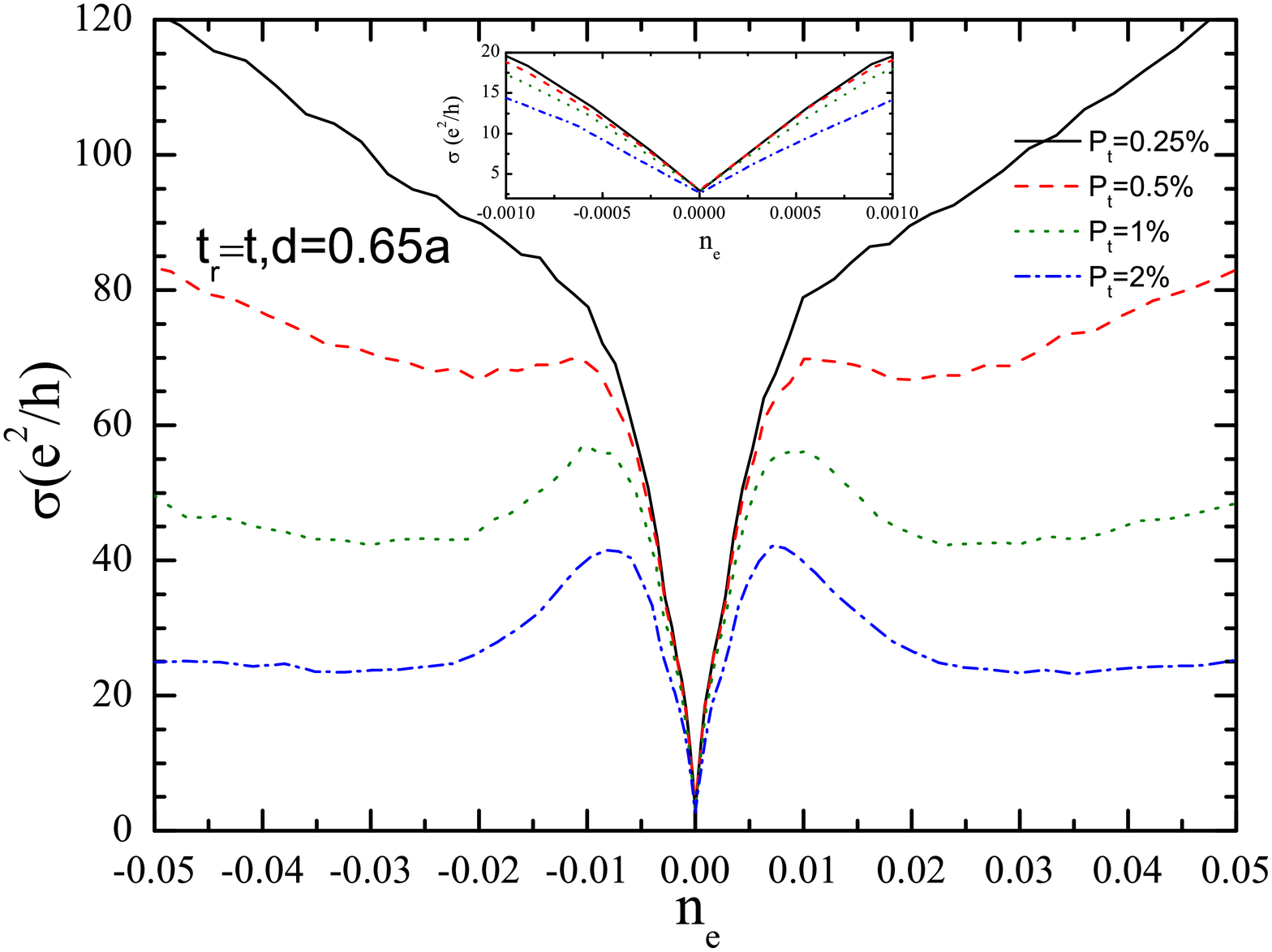}
\includegraphics[width=7.5cm]{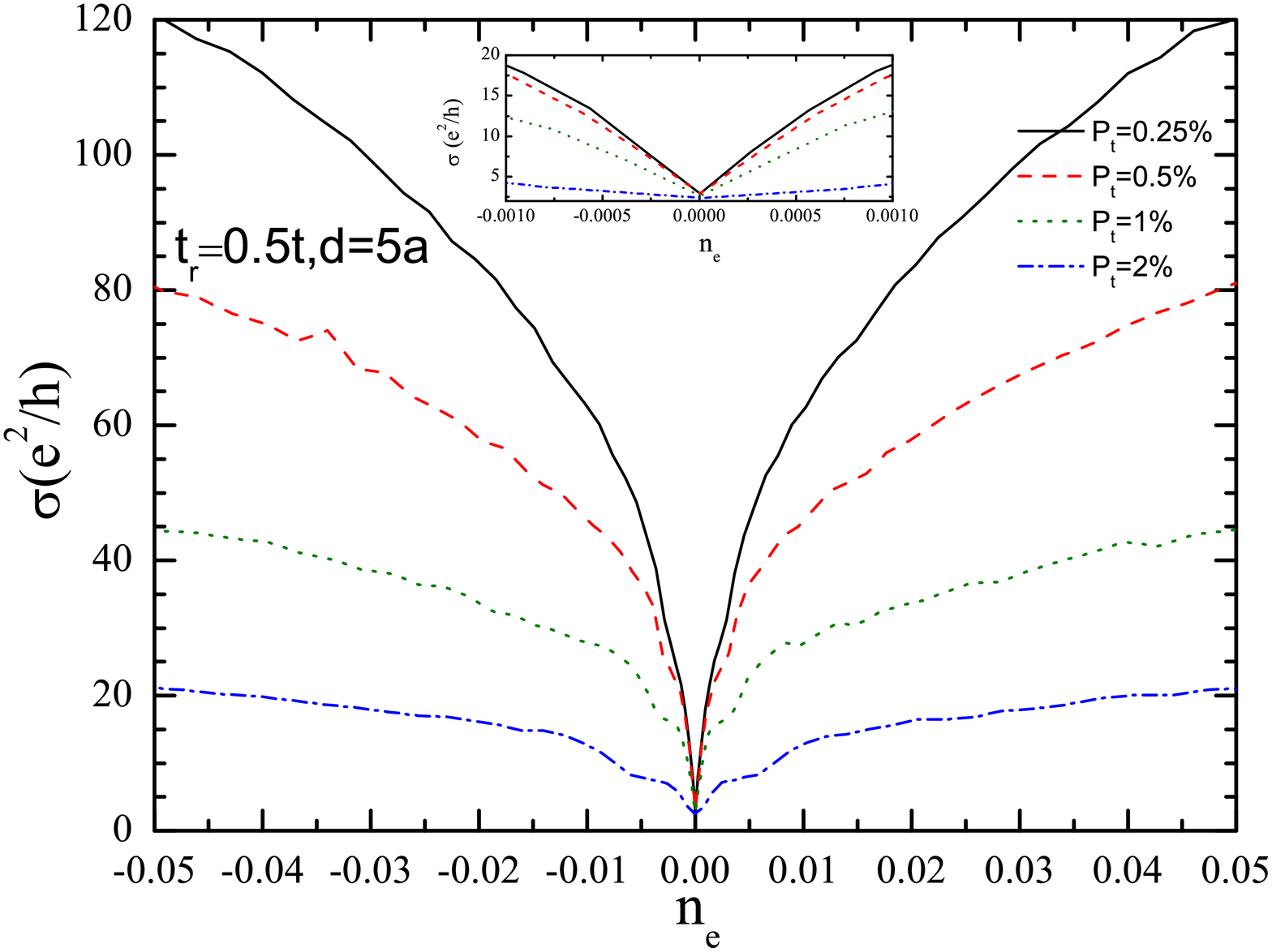}
}
\end{center}
\caption{(Colour online) DOS and conductivity of bilayer graphene ($\protect%
\gamma _{1}=\protect\gamma _{3}=0.1t$) with short-range ($\Delta _{t}=t,$ $%
d_{t}=0.65a$) or long-range ($\Delta _{t}=0.5t,$ $d_{t}=5a$) Gaussian
hopping. Each layer contains $4096\times 4096$ carbon atoms.}
\label{trgaussian}
\end{figure*}

\begin{figure*}[t]
\begin{center}
\mbox{
\includegraphics[width=7.5cm]{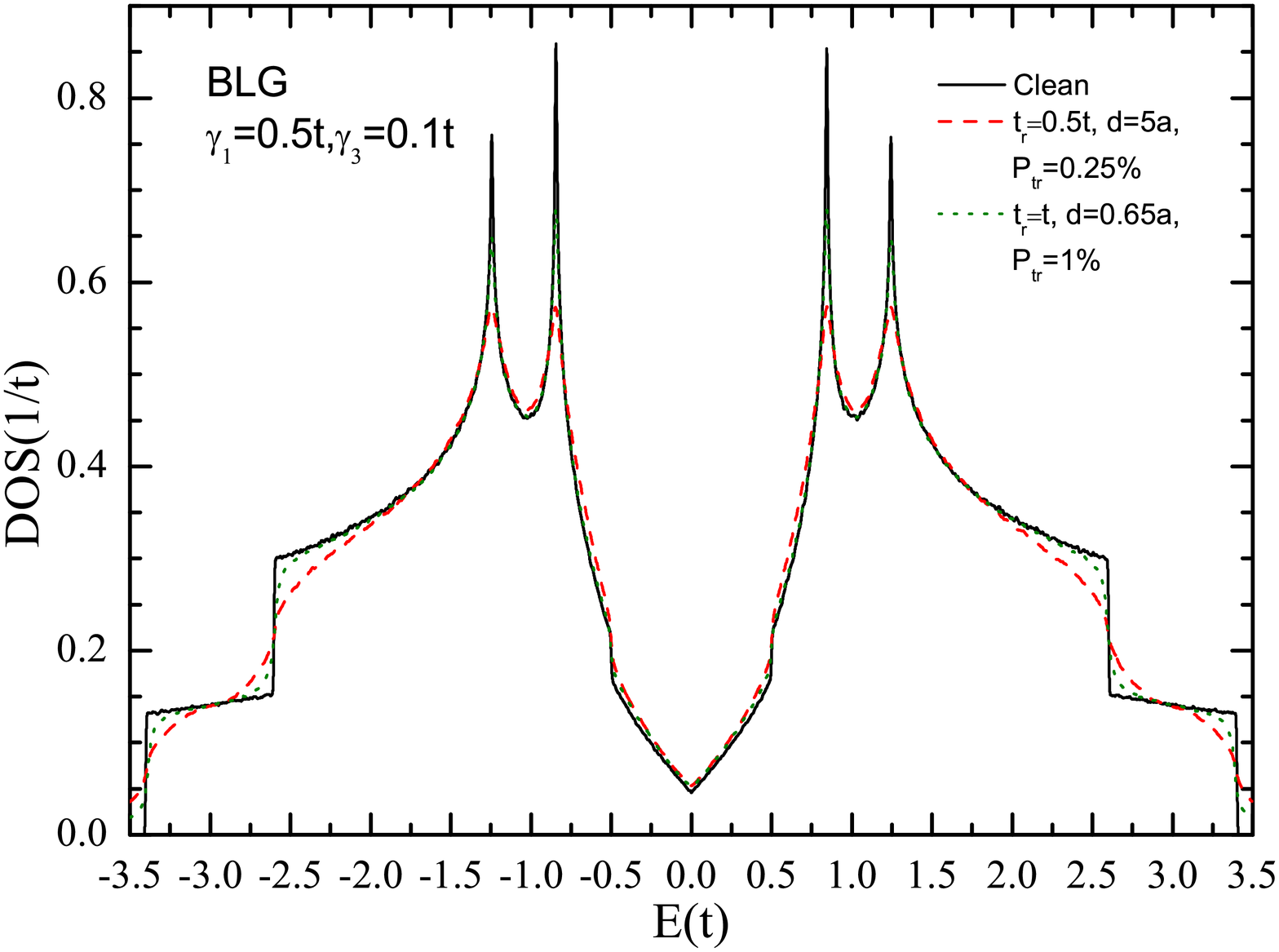}
\includegraphics[width=7.5cm]{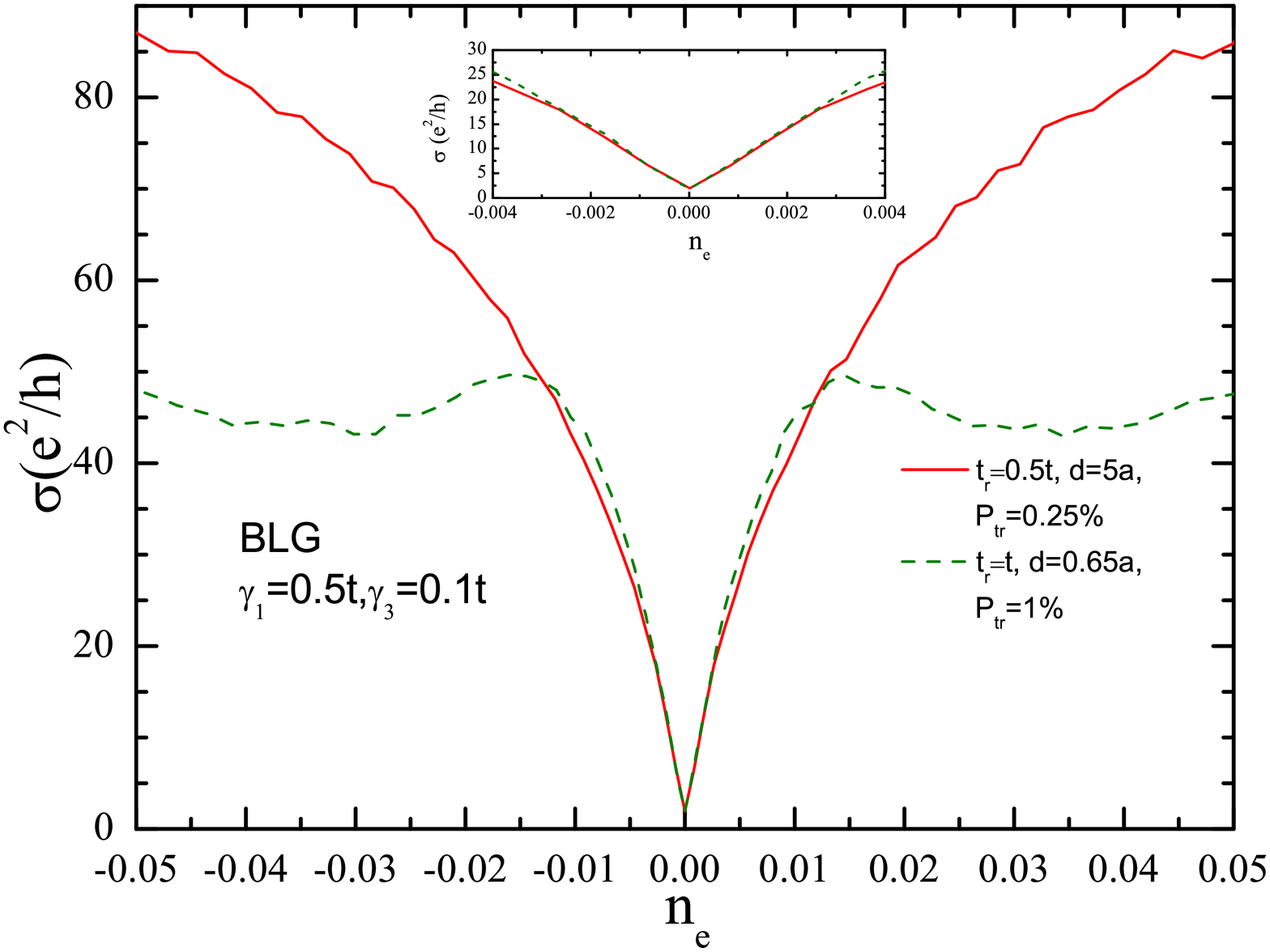}
}
\mbox{
\includegraphics[width=7.5cm]{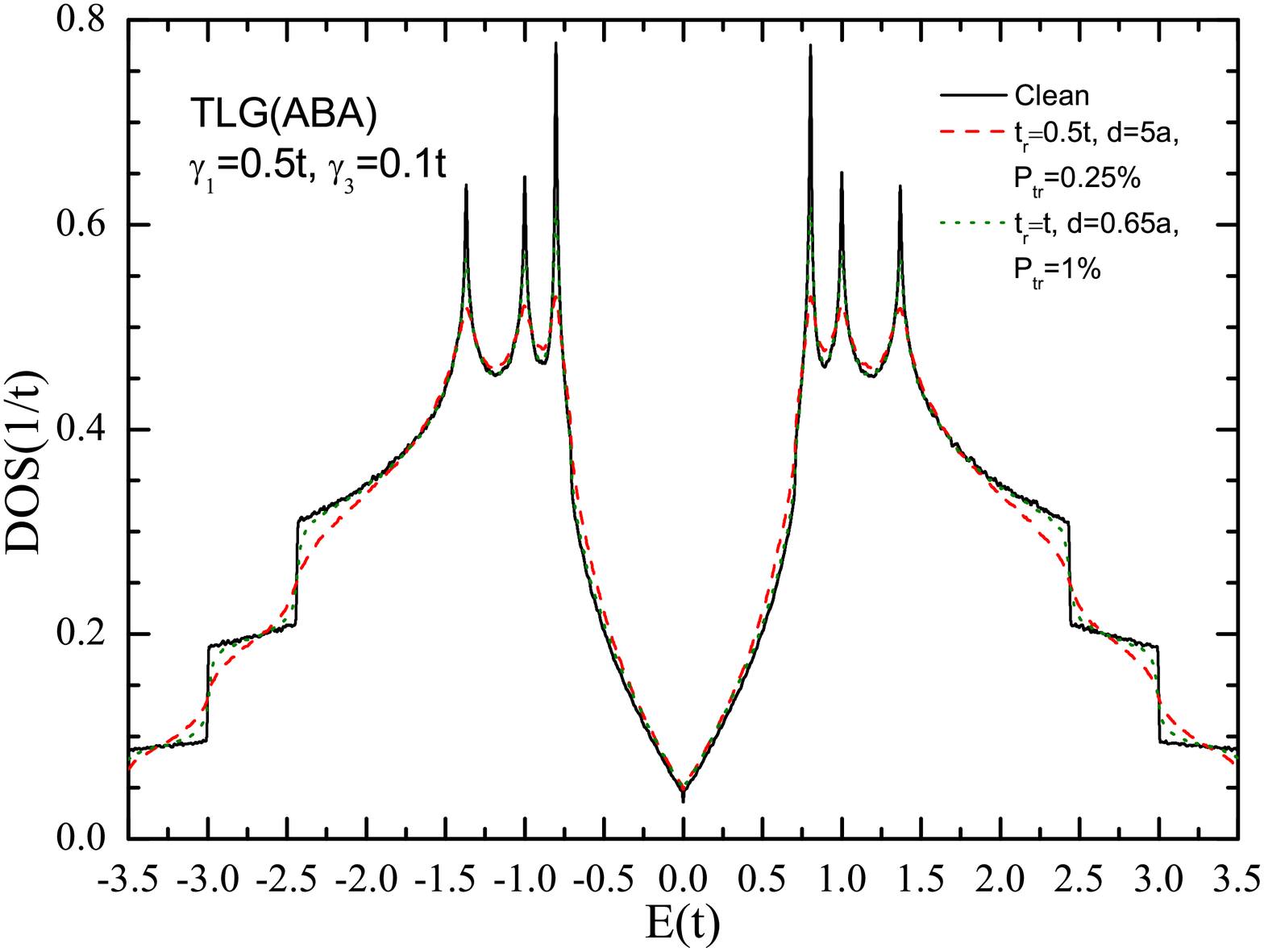}
\includegraphics[width=7.5cm]{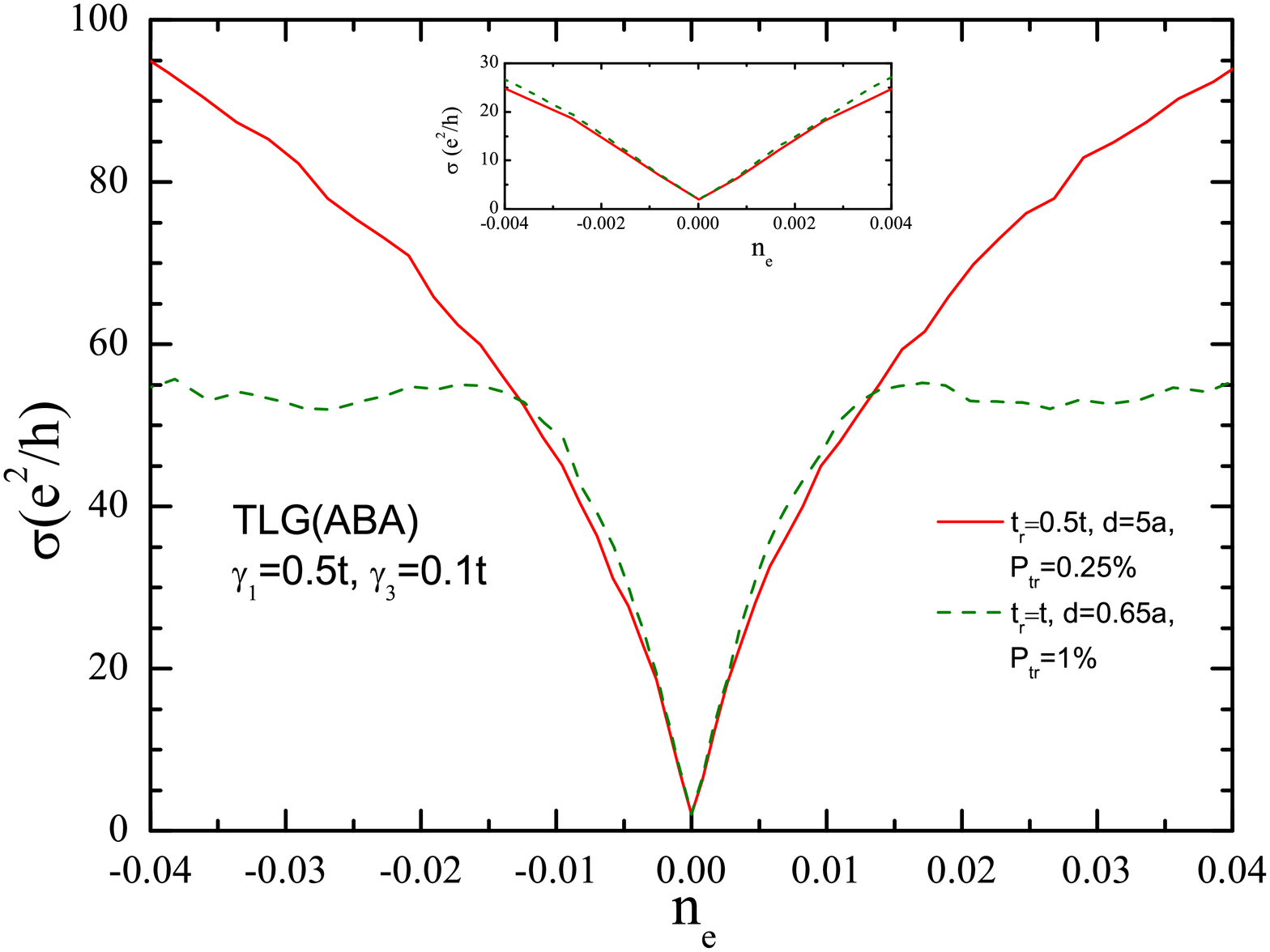}
}
\mbox{
\includegraphics[width=7.5cm]{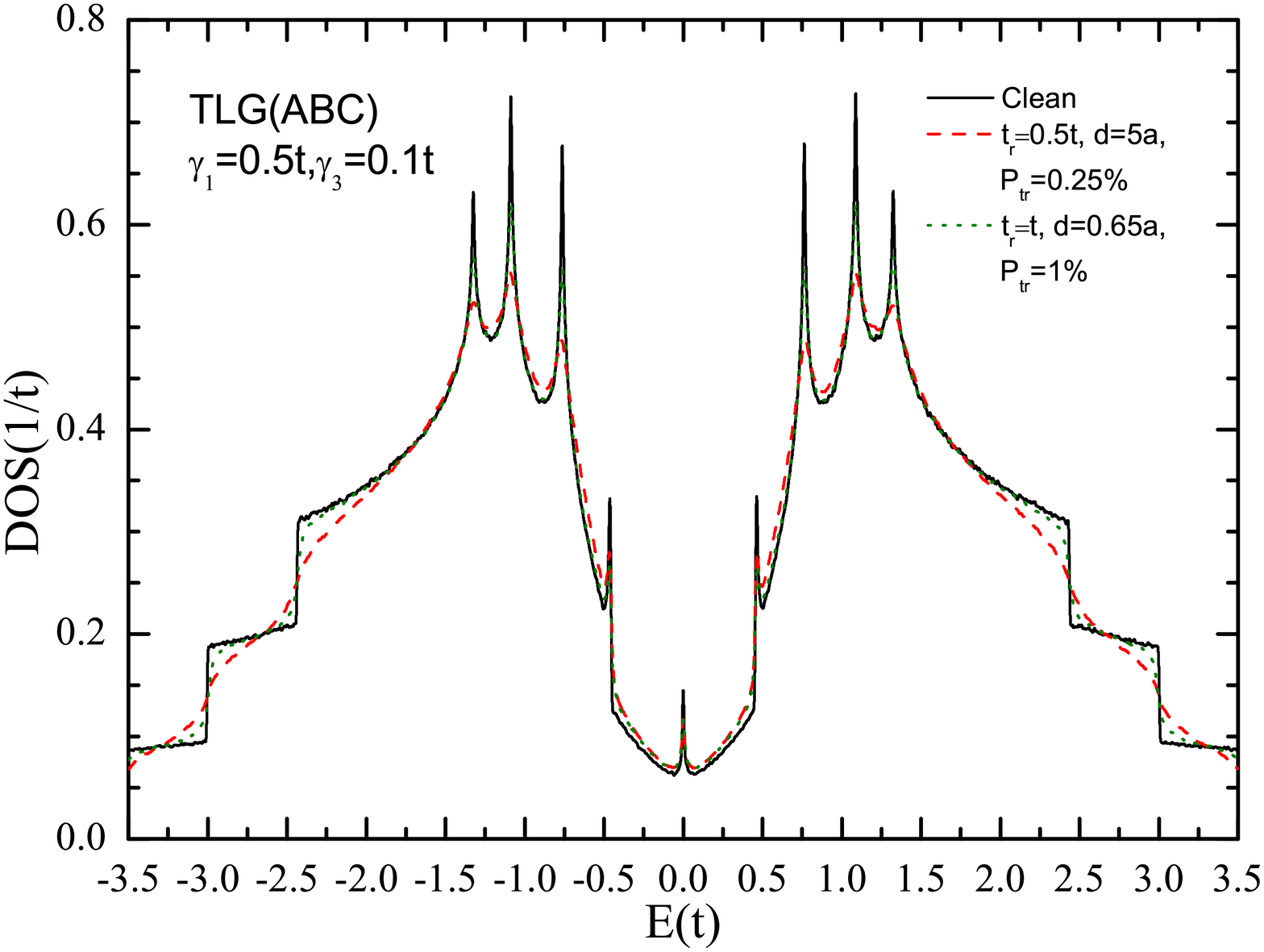}
\includegraphics[width=7.5cm]{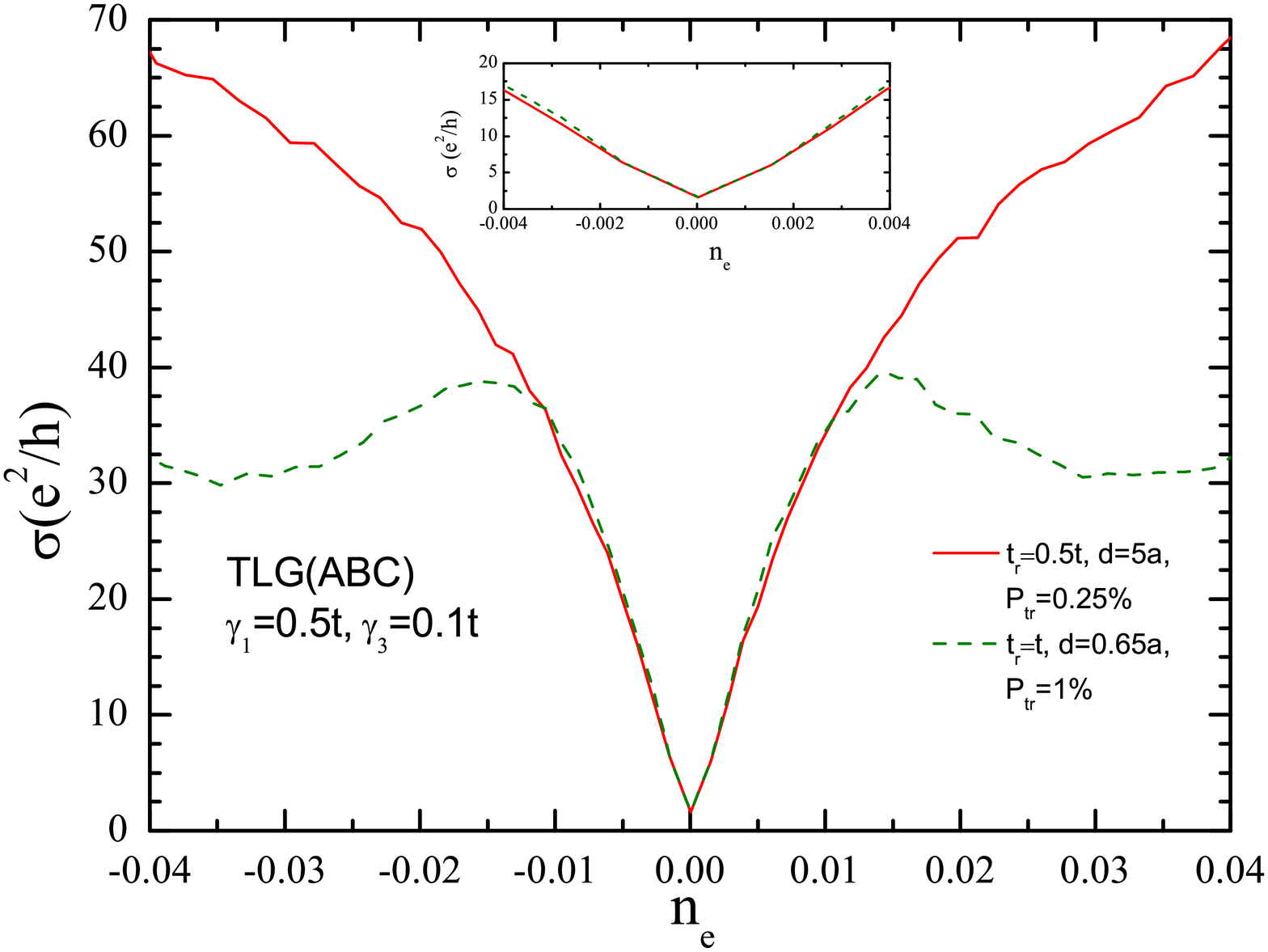}
}
\end{center}
\caption{(Colour online) DOS and conductivity of bilayer graphene ($\protect%
\gamma _{1}=\protect\gamma _{3}=0.1t$) with short-range ($\Delta _{t}=t,$ $%
d_{t}=0.65a$) or long-range ($\Delta _{t}=0.5t,$ $d_{t}=5a$) Gaussian
hopping. Each layer contains $4096\times 4096$ carbon atoms.}
\label{trgaussianc}
\end{figure*}

Previous studies of the vacancies in BLG focused mainly on the properties of
the local density of states (LDOS) around a single or a pair of vacancies,
and it was shown that the LDOS in the neighboring lattice sites of the
impurity site is normally enhanced, depending on the lattice site (A or B
sublattices) of the vacancy \cite{Wang2007,Dahal2008}. Recently, a new type
of zeromodel state in BLG is found in Ref. \onlinecite{Castro2010b}, in the
absence of a gap it is quasilocalized in one of the layers and delocalized
in the other, and in the presence of a gap it becomes fully localized inside
the gap. These observations are different from SLG, where the impurity state
is insensitive to the position of vacancies. The differences in the spectrum
of LDOS around the vacancies in SLG and BLG lead to different
electron-density (Fermi energy) dependence of the conductivity. As the
vacancies and resonant impurities have similar effects on the electronic
structure and transport properties in SLG \cite{Wehling2010,Yuan2010}, it
suggests that their contributions to the bilayer and trilayer graphene
should also be comparable. We consider here the results for the vacancies in
the range that the Boltzmann approach is applicable. In Fig. \ref{dcvacancy}%
, we show the numerical results of the DOS and conductivities of SLG, BLG
and TLG with fixed concentration of vacancies ($n_{x}=0.5\%$). The
parameters of the interlayer coupling are $\gamma _{1}=0.5t$ and $\gamma
_{3}=0.1t$. These results are directly comparable with the results of the
same concentration of resonant impurities represented in Fig. \ref%
{ximpslbltri}, and demonstrate similar density-dependence of the
conductivities, just as we expected. For conciseness, we do not discuss
these vacancies as their effect on the transport properties of graphene are
quite similar to those of the resonant impurities.

\section{Gaussian Potential}

The impurities in the Hamiltonian of Eq.~(\ref{Hamiltonian}) are represented
by random on-site potentials. Short-rang and long-range Gaussian potentials
are given by
\begin{equation}
v_{i}=\sum_{k=1}^{N_{imp}^{v}}U_{k}\exp \left( -\frac{\left\vert \mathbf{r}%
_{i}-\mathbf{r}_{k}\right\vert ^{2}}{2d^{2}}\right) ,  \label{vgaussian}
\end{equation}%
where $N_{imp}^{v}$ is the number of the Gaussian centers, which are chosen
randomly distributed on the carbon atoms, $U_{k}$ is uniformly random in the
range $[-\Delta ,\Delta ]$ and $d$ is interpreted as the effective potential
radius. The typical values of $d$ used in our model are $d=0.65a$ and $5a$
for short- and long-range Gaussian potential, respectively. Here $a$ is the
carbon-carbon distance in the monolayer graphene. The value of $N_{imp}^{v}$
is characterized by the value $P_{v}=N_{imp}^{v}/N$, where $N$ is the total
number of carbon atoms of the sample. A typic contour plot of the on-site
potentials in the central part of a graphene layer with short- or long-range
Gaussian potential is shown in Fig. \ref{contourpotential}. The sum in Eq.~(%
\ref{vgaussian}) is limited to the sites in the same layer, i.e., we do not
consider the overlapping of the Gaussian distribution in different layers.

Numerical results of the density of states and dc conductivities of BLG ($%
\gamma _{1}=\gamma _{3}=0.1t$) with short- ($\Delta =3t,$ $d=0.65a$) and
long-range ($\Delta =1t,$ $d=5a$) Gaussian potentials are shown in Fig. \ref%
{gaussian}. Similar to the case of resonant impurities, the singularities in
the spectrum are also suppressed in the presence of random potentials, and
the conductivity as a function of charge density follows a sublinear
dependence. The difference is that there is no impurity band around the
neutrality point (see the DOS in Fig. \ref{gaussian}). This leads to totally
different transport properties: no plateau around the Dirac point in the
conductivity vs. $n_{e}$ curves.

Similar to the case of resonant impurities, the regime of parabolic band in
BLG expands by increasing $\gamma _{1}$ from $0.1t$ to $0.5t$, the results
being shown in Fig. \ref{gaussian_gemma1}. Now the difference of transport
properties in BLG with short- and long-range Gaussian potentials are more
significant within the parabolic band: the density-dependence of
conductivity is sublinear in the case of short-range, but linear in the case
of long-range potentials. Actually, these sublinear and linear dependencies
are also observed in TLG, independent on the stacking sequence (see Fig. \ref%
{gaussian_gemma1}).

The same value of the minimum conductivity ($\sigma _{\min }\approx 2e^{2}/h$%
) at the charge neutrality point is observed for both BLG and TLG with $%
\gamma _{1}=0.5t$. As we discussed in the case of resonant impurities, the
adoption of larger $\gamma _{1}$ is equivalent to the use of smaller
disorder, and therefore our results indicate that the minimum conductivity
in order of $\sigma _{\min }\approx 2e^{2}/h$ is common in BLG and TLG with
small concentration of random Gaussian potentials. These numerical results
are consistent with the analytical result for BLG in Ref. %
\onlinecite{Katsnelson_bilayer}.

\section{Gaussian Hopping}

The origin of disorder in the nearest neighbor coupling could be
substitutional impurities like N or B instead of C, or distortions of
graphene sheet. To be specific, we introduce the disorder in the hopping by
a Gaussian distribution in a similar way as random Gaussian potential,
namely, the distribution of the nearest neighbor hopping parameters reads%
\begin{equation}
t_{ij}=t+\sum_{k=1}^{N_{imp}^{t}}T_{k}\exp \left( -\frac{\left\vert \mathbf{r%
}_{i}+\mathbf{r}_{j}-2\mathbf{r}_{k}\right\vert ^{2}}{8d_{t}^{2}}\right) ,
\label{tgaussian}
\end{equation}%
where $N_{imp}^{t}$ is the number of the Gaussian centers, $T_{k}$ is
uniformly random in the range $[-\Delta _{t},\Delta _{t}]$ and $d_{t}$ is
interpreted as the effective screening length. Similarly, the typical values
of $d_{t}$ are the same as for the Gaussian potential, i.e., $d_{t}=0.65a$
and $5a$ for short- and long-range Gaussian random hopping, respectively,
and the values of $N_{imp}^{t}$ are characterized by the value $%
P_{t}=N_{imp}^{t}/N$. Similar as in Eq.~(\ref{vgaussian}), the sum in Eq.~(%
\ref{tgaussian}) does not include the overlapping of the Gaussian
distribution in different layers.

Like in the case of Gaussian potentials, the presence of random Gaussian
hopping in BLG and TLG also suppresses the Van Hove singularities in the
spectrum, but does not introduce a new impurity band (midgap states) and
there is also no plateau in the conductivity vs. electron density curves
(see Fig. \ref{trgaussian} and \ref{trgaussianc}). The unique feature
characteristic for the presence of random Gaussian hopping is that in the
region near the neutrality point, the conductivity is always linearly
dependent on the electron density, with no influence from the concentration
of Gaussian centers (different $P_{t}$ in Fig. \ref{trgaussian}), range of
Gaussian coupling ($d_{t}=0.65a$ or $5a$), strength of the interlayer
coupling ($\gamma _{1}=0.1t$ in Fig. \ref{trgaussian} and $0.5t$ in Fig. \ref%
{trgaussianc}), number of layers (bilayer or trilayer) and stacking sequence
(ABA or ABC in TLG). The differences of short- or long-range cases are only
obvious in the energy region far from the neutrality point (high
concentration of charge density): the increase of conductivity as a function
of charge density is monotonic only for the long-range disorder.
Furthermore, like in the case of random Gaussian potential, a common minimum
conductivity in the order of $2e^{2}/h$ on charge neutrality point is also
observed for both BLG and TLG.

\section{Discussion and Conclusions}

We have presented a detailed numerical study of the electronic transport
properties of bilayer and trilayer graphene within the framework of a
noninteracting tight-binding model. Various realistic types of disorder are
considered, such as resonant impurities, vacancies, random Gaussian on-site
potentials, and random Gaussian hopping between nearest carbon atoms. Our
results give a consistent picture of the electronic structure and transport
properties of bilayer and trilayer graphene in a broad range of
concentration of impurities or other sources of disorder. Linear or
sublinear electron-density dependent conductivity at high enough density is
observed, depending on the type and strength of the disorder and the
stacking sequence. The minimum conductivity $\sigma _{\min }\approx 2e^{2}/h
$ (per layer) on charge neutrality point is common for BLG and TLG,
independent of the type of the impurities, but the plateau of minimum
conductivity around the neutrality point is unique when resonant impurities
or vacancies are present.

In the presence of resonant impurities or vacancies, the dependence of the
conductivity as a function of electron density is affected by the relevant
width of the impurity band and the band created by the interlayer hopping.
Using BLG with vacancies as an example: introducing $n_{p}\equiv n_{e}\left(
\gamma _{1}\right) =\int_{0}^{\gamma _{1}}\rho \left( \varepsilon \right)
d\varepsilon $\ as the density of electrons on the boundary of the parabolic
band, and considering the case that the concentration of vacancies ($n_{x}$)
is smaller than $n_{p}$, i.e., the impurity band is within the region of the
parabolic conduction band, there are three regions of electron-density
dependence of the conductivity:

(i) $\left\vert n_{e}\right\vert \leq n_{x}$, a central minimum conductivity
plateau ($2e^{2}/h$\ per layer) with width equals to $2n_{x}$;

(ii) $n_{x}<\left\vert n_{e}\right\vert <n_{p}$, linear dependence, as
predicted by the analytical treatment using the Boltzmann equation for
parabolic spectrum \cite{Katsnelson_bilayer};

(iii) $n_{x}\geq n_{p}$, sublinear dependence, as the effects of the
interlayer hopping are negligible in this region and one should expect a
behavior of the conductivity similar to that of SLG.

On the opposite case $n_{x}\geq n_{p}$, region (ii) simply disappears and
therefore we can only observe the minimum conductivity plateau and sublinear
dependence on the high concentration of electron densities. Actually, the
sublinear dependence beyond the parabolic band is a general property of SLG,
BLG and MLG with large enough concentration of resonant impurities or
vacancies, independent on the number of layers and the stacking sequence.

In the presence of random Gaussian on-site potentials, the electron-density
dependences of conductivity of BLG or TLG are sublinear and linear in the
low concentration of charges, for short- and long-range disorders,
respectively but are always sublinear in the high concentration. On the
other hand, in the case of random Gaussian carbon--carbon couplings, the
density-dependence of conductivity in the region close to the neutrality
point is more simple: there is only a linear dependence, with no effect of
the strength and range of disorder, the number of layers and stacking
sequence.

Note added: After this paper was submitted, a paper which also
discusses the effect of resonant scatterers on the dc conductivity
of single-layer and bilayer graphene appeared~\cite{Ferreira2010},
with results that are consistent with ours.

\section{Acknowledgement}

The support by the Stichting Fundamenteel Onderzoek der Materie (FOM) and
the Netherlands National Computing Facilities foundation (NCF) are
acknowledged.

\end{document}